\documentclass[paper=a4,fontsize=11pt,twoside]{article}
\usepackage{authblk} 

\title{Pragmatic hypotheses in \\ the evolution of Science}

\author[1]{Lu\'is Gustavo Esteves}
\author[2]{\mbox{} Rafael Izbicki}  
\author[2]{ \authorcr Rafael Bassi Stern}
\author[1]{\mbox{} Julio Michael Stern\thanks{Authors' e-mails:  \ {\tt jstern@ime.usp.br, jmstern@hotmail.com} (J.M.S., corresponding author); \ 
{\tt lesteves@ime.usp.br} (L.G.E.); \  {\tt rafaelizbicki@gmail.com} (R.I.); \  {\tt rbstern@gmail.com} (R.B.S.). } } 

\affil[1]{University of S\~{a}o Paulo}
\affil[2]{Federal University of S\~{a}o Carlos}

\date{December 22, 2018}

\usepackage[T1]{fontenc}
\usepackage{nameref}
\usepackage[colorlinks=true, urlcolor=blue, citecolor=blue, linkcolor=blue]{hyperref}
\usepackage{amsfonts, amsmath, amssymb, amsthm, thmtools}
\usepackage{mathrsfs}
\usepackage{cleveref}

\declaretheorem[name=Theorem, refname={Theorem, Theorems}, Refname={Theorem, Theorems}, parent=section]{theorem}

\declaretheorem[name=Corollary, refname={Corollary, Corollaries}, Refname={Corollary, Corollaries}, sibling=theorem]{corollary}
\declaretheorem[name=Definition, refname={Definition, Definitions}, Refname={Definition, Definitions}, sibling=theorem, style=definition]{definition}
\declaretheorem[name=Example, refname={Example, Examples}, Refname={Example, Examples}, sibling=theorem, style=definition]{example}

\crefname{section}{section}{sections}
\Crefname{section}{Section}{Sections}
\crefname{table}{table}{tables}
\Crefname{table}{Table}{Tables}

\usepackage{color, enumitem, float, fourier, layout, multirow, setspace, verbatim}
\setlist[enumerate]{leftmargin=*}
\usepackage[colorinlistoftodos]{todonotes}
\usepackage{algorithm2e, algorithmicx, listings}
\usepackage{caption, subcaption}
\usepackage{natbib}
\usepackage{enumitem}
\usepackage{graphicx}
\graphicspath{{figures/}{../figures/}}

\makeatletter
\def\widebreve{\mathpalette\wide@breve}
\def\wide@breve#1#2{\sbox\z@{$#1#2$}%
	\mathop{\vbox{\m@th\ialign{##\crcr
				\kern0.08em\brevefill#1{0.8\wd\z@}\crcr\noalign{\nointerlineskip}%
				$\hss#1#2\hss$\crcr}}}\limits}
\def\brevefill#1#2{$\m@th\sbox\tw@{$#1($}%
	\hss\resizebox{#2}{\wd\tw@}{\rotatebox[origin=c]{90}{\upshape(}}\hss$}
\makeatletter

\def\half{\frac{1}{2}}
\def\btheta{\boldsymbol{\theta}}

\def\I{{\mathbb I}}

\def\Z{{\vec{Z}}}
\def\z{{\vec{z}}}

\def\E{{\textbf{E}}}

\def\P{{\mathbb P}}
\def\S{{\mathbf{S}}}
\def\V{{\mathbb V}}
\def\Re{{\mathbb R}}

\def\hT{{{\hat{\theta}}}}
\def\hZ{{{\hat{\Z}}}}

\def\sZ{\mathcal{Z}}
\def\sX{\mathcal{X}}

\def\t0{{\theta_0}}
\def\ts{{\theta^*}}
\def\tr{\mbox{tr}}

\renewcommand{\vec}[1]{\mathbf{#1}}

\usepackage{pgfplots}
\pgfplotsset{compat=1.14}

\newcommand{\gfbstfig}{
 \begin{tikzpicture}[thick,scale=0.75, every node/.style={scale=0.9}]
  \draw (-6,2) circle (1.5);
  \draw [fill=lightgray] (-6,2) circle (0.8);
  \draw (-8,0) -- (-4,0) -- (-4,4) -- (-8,4) -- (-8,0);;
  \node at (-6,2) {{\large$R(x)$}}; 
  \node at (-6,0.8) {{\large$H_0$}};
  \node at (-7.5,.50) {{\large$H_0^c$}};
  \node at (-6,4.25) {{\large 	$\phi^R(x)=0$}};
			
  \draw (-1.4,2.4) circle (1.1);
  \draw [fill=lightgray] (0.2,0.9) circle (0.7);
  \draw (-3,0) -- (1,0) -- (1,4) -- (-3,4) -- (-3,0);;
  \node at (0.2,0.9) {{\large$R(x)$}}; 
  \node at (-1.4,2.4) {{\large$H_0$}};
  \node at (-2.5,.50) {{\large$H_0^c$}}; 
  \node at (-1,4.25) {{\large $\phi^R(x)=1$}};	
			
  \draw [fill=lightgray] (5.2,1.9) circle (0.7); 
  \draw (4.4,2.4) circle (1.1);
  \draw (2,0) -- (6,0) -- (6,4) -- (2,4) -- (2,0);;
  \node at (5.2,1.9) {{\large$R(x)$}}; 
  \node at (4.2,2.5) {{\large$H_0$}};
  \node at (2.5,.50) {{\large$H_0^c$}};
  \node at (4,4.25) {{\large $\phi^R(x)=1/2	$}};	
 \end{tikzpicture} \\ 
}

\def\ts{{\theta^*}}
\def\tz{{\theta_0}}

\begin{document}

\maketitle

\mbox{} 

\vfill 

\pagebreak 

\begin{abstract}
 This paper introduces pragmatic hypotheses and
 relates this concept to the spiral of scientific evolution.
 Previous works determined a characterization of
 logically consistent statistical hypothesis tests and
 showed that the modal operators obtained from this test
 can be represented in the hexagon of oppositions.
 However, despite the importance of precise hypothesis in science,
 they cannot be accepted by logically consistent tests.
 Here, we show that this dilemma can be
 overcome by the use of pragmatic versions of
 precise hypotheses. These pragmatic versions allow
 a level of imprecision in the hypothesis that is
 small relative to other experimental conditions.
 The introduction of pragmatic hypotheses allows
 the evolution of scientific theories based on
 statistical hypothesis testing to be interpreted
 using the narratological structure of hexagonal spirals,
 as defined by Pierre Gallais.  
\end{abstract}

\section{Introduction}
\label{sec:intro} 

Standard hypothesis tests can lead to
immediate logical incoherence, which
makes their conclusions hard to interpret.
This incoherence is a result of
such tests having only two possible outcomes.
Indeed, \citet{Izbicki2015} shows that
there exists no two-valued test that
satisfies desirable statistical properties and
that is also logically coherent.

\begin{figure}[ht] 
 \centering
 \includegraphics*[scale=0.25]{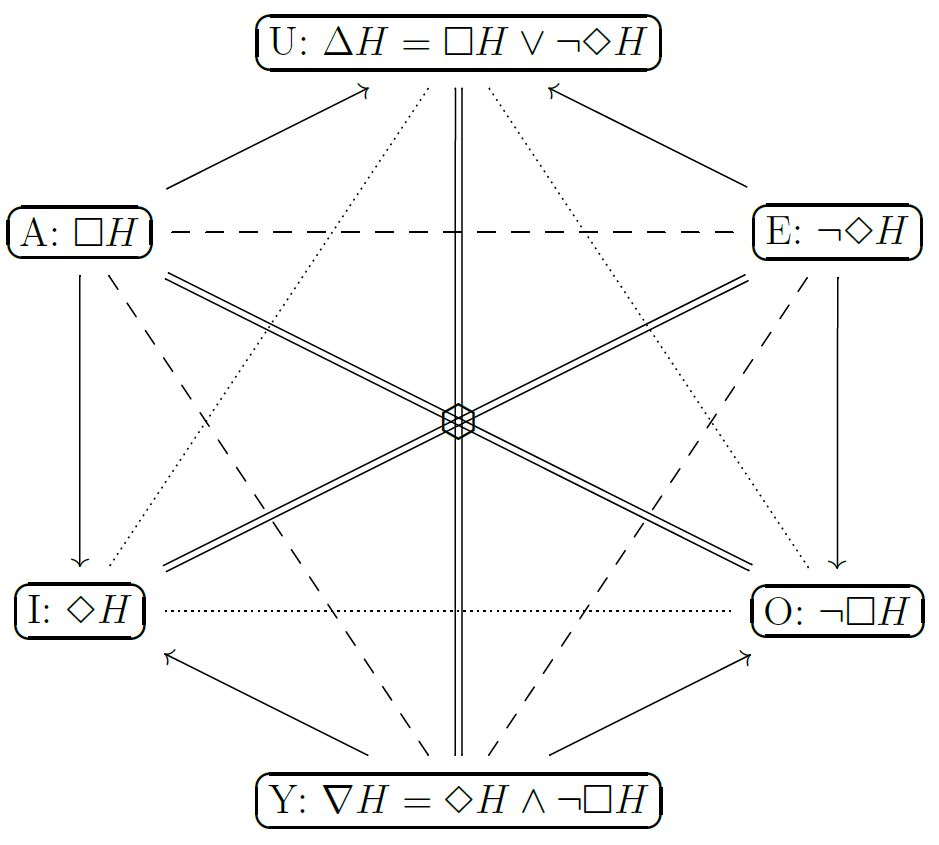}
 \caption{Hexagons of Opposition for Statistical Modalities.}
 \label{fig:hexagon}
\end{figure}

In order to overcome such an impossibility result,
\citet{Esteves2016} proposes agnostic hypothesis tests,
which have three possible outputs:
(A) accept the hypothesis, say $H$, (E) reject $H$,
or (Y) remain agnostic about $H$.
These tests can be made logically coherent while
preserving desirable statistical properties.
For instance, both conditions are satisfied by
the Generalized Full Bayesian Significance Test (GFBST).
Furthermore, \citet{Stern2017b} shows that
the GFBST's modal operators and their respective negations can
be represented by vertices of the hexagon of oppositions
\citep{Blanche1966,Beziau2012,Beziau2015,Carnielli2008,Dubois1982,Dubois2012},
which is depicted in \cref{fig:hexagon}. 

This paper complements the above static representation with
an analysis of the GFBST in the dynamic evolution of 
scientific theories.
The analysis is based on the
metaphor of evolutive hexagonal spirals
\citep{Gallais1974,Gallais1982}, in which
the logical modalities associated to
scientific theories change over time,
as in \cref{fig:spiral}.
Our key point in this paradigm is
reconciling two apparently contradictory facts.
On the one hand, precise or sharp hypotheses, that is, hypotheses that have \emph{a priori} zero probability are
central in scientific theories \citep{Stern2011b,Stern2017a}.
On the other hand, the GFBST never
accepts (A) precise hypotheses.
These observations lead to the apparent paradox that,
if the GFBST were used to test scientific theories, then
the acceptance step in the spiral of
scientific theories would be forfeited.

In order to overcome this paradox,
we propose the concept of a \textit{pragmatic hypothesis}
associated to a precise hypothesis.
While precise hypothesis are commonly 
obtained from mathematical theories used in areas of science and technology
\citep{Stern2011b, Stern2017a},
the associated pragmatic hypothesis is
an imprecise hypothesis which is
sufficiently good from the practical purpose of
an end-user of the theories.
For instance, Newtonian theory assumes
a gravitational force of magnitude given by
the equation $F=G\,m_1\,m_2\,d^{-2}$, where
the gravitational constant $G$ has a precise value.  
However, the current CODATA (Committee on Data for
Science and Technology) value for
the gravitational constant is 
$G=6.67408(31)\times 10^{-11} {\mbox{m}^{3}\ \mbox{kg}^{-1} \
\mbox{s}^{-2}}$, which includes
a standard deviation for the last signicant digits, $408\pm31$. 
Hence, it may be reasonable for a given end-user to
assume that the theoretical form of the last equation is
exact, but that, pragmatically,
the constant $G$ can only be known up to
a chosen precision. As a result,
one might wish to test an imprecise hypothesis
associated to the scientific hypothesis of interest
\citep{Degroot2012,Berger2013}

This article advocates for the conceptual distinction between
a precise scientific theory and an associated pragmatic hypotheses.
The alternate use of precise and pragmatic versions of
corresponding statistical hypotheses enables
the GFBST to (pragmatically) accept
scientific hypotheses. Moreover,
this alternate use allows the GFBST to
track the evolution of scientific theories,
as interpreted in the context of
Gallais' hexagonal spirals.
 
Our main goal in this paper is to
formalize testing procedures for
a theory taking into consideration the level of
precision that is appropriate for a given end-user.
In order to handle this problem,
we consider the end-user's predictions about
an experiment of his interest.
The variation in these predictions can
be explained by a combination of
the level of imprecision in the theory and
by properties of the end-user's experiment.
For instance, the latter source of variation is
influenced by properties of the equipment,
including precision, accuracy and
resolution of measuring devices
\citep{Bucher2012, Czichos2011}, and 
also error bounds for fundamental constants and
calibration factors \citep{Cohen1957, Cohen1957b, Leblond1977, Pakkan1995, Akman1996, Wainwright2002, Bishop2006, Iordanov2010, Gelman2014}.
We propose to choose
a pragmatic hypothesis in such a way that
the imprecision in the end-user's predictions is
mostly due to his experimental conditions and
not due to the level of imprecision in
the theory that he uses.
 
In order to develop this argument,
\cref{sec:gallais} first adapts
Gallais's metaphor of hexagonal spirals to
the evolution of science. Next,
\cref{sec:predictive}, proposes
three methods of decomposing
the variability in an end-user's predictions into
the level of precision of the theory he uses and
his experimental conditions.
\Cref{sec:simple,sec:composite} use
these decompositions in order to
build pragmatic hypothesis.
They build pragmatic hypotheses for
simple hypotheses and then
prove that there exists a single way of
extending this construction to
composite hypotheses while
preserving logical coherence in
simultaneous hypothesis testing.
This methodology is illustrated in
\cref{sec:applications}.
All proofs are found in \cref{sec:proof}.

\section{Gallais' hexagonal spirals and
the evolution of science} 
\label{sec:gallais}

\begin{figure} 
 \centering
 \includegraphics*[height = 3.1in, width = 3.3in, angle = 0,
 trim = 0mm 0mm 0mm 0mm, clip]{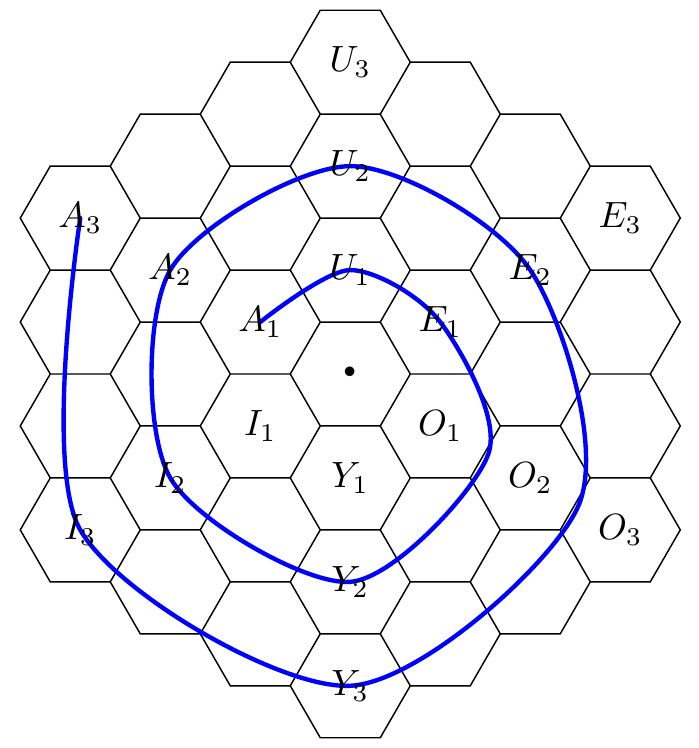} \\ 
 \caption{Gallais' evolutionary spiral.} 
 \label{fig:spiral} 
\end{figure}

Following a well-established tradition in structural semantics and
narratology \citep{Greimas1983,Propp2000},
\citet{Gallais1974} proposes that
many classical medieval tales follow
the same organizational pattern. More precisely,
these narratives exhibit an
underlying \textit{intellectual structure} and
are organized according to an underlying archetypal format or
prototypical pattern. This pattern includes both
static and a dynamical aspects.
From a static perspective,
the logical structure of the narrative is such that
each arch is represented by a vertex of
the \textit{hexagon of oppositions} \citep{Blanche1966}.
The static hexagon of oppositions is
depicted in \cref{fig:hexagon} and
represents in each vertex a modal operator among
necessity ($\square$), possibility ($\Diamond$),
contingency ($\Delta$) and their negations ($\lnot$).
These modal operators are structured according to
three axes of opposition $(===\,)$,
a triangle of contrariety $(- - -\,)$,
another triangle of sub-contrariety, $(\cdots\,)$, and
several edges of subalteration $(\longrightarrow$\,).
From a dynamical perspective, the temporal evolution of
the narrative follows a spiral (\cref{fig:spiral}) that
unwinds (\textit{se d\'{e}roule}) around
concentric and expanding hexagons of opposition
\citep{Gallais1974,Gallais1982}.

Since also the evolution of science can be conceived as following a spiral pattern \citep{Stern2014b}, its analysis can
benefit from the structure in
\citep{Gallais1974, Gallais1982}.
From a static perspective,
the logical modalities induced by
agnostic hypothesis tests \citep{Stern2017b} can be
represented in the hexagon of oppositions.
From a dynamic perspective, scientific theories
evolve as a spiral which
unwinds around the following states:
\begin{itemize}[leftmargin=*]
 \item \textbf{$A_1$- Extant thesis:} 
 This vertex represents a standing paradigm, an \textbf{accepted} theory
using well-known formalisms and familiar concepts, relying on accredited experimental
means and methods, etc. In fact, the concepts of a current paradigm may become
so familiar and look so natural that they become part of a reified ontology. 
 That is, there is a perceived correspondence between
 concepts of the theory and \textit{dinge-an-sich} 
  (things-in-themselves) as
 seen in nature \citep{Stern2011a,Stern2014b}.
 
 \item \textbf{$U_1$- Analysis:} 
 This vertex represents the moment when some hypotheses of the
standing theory are put in \textbf{question}. At this moment, possible alternatives to the standing
hypotheses may still be only vaguely defined
  
 \item \textbf{$E_1$- Antithesis:}
 This vertex represents the moment when some laws of the standing theory have to be \textbf{rejected}. Such a rejection of old laws may put in question the entire world-view of the current paradigm, opening the way for revolutionary ideas, as described in the next vertex.
  
 \item \textbf{$O_2$-  Apothesis/ Prosthesis:} 
This vertex is the locus of revolutionary freedom. Alternative
models are considered, and specific (precise) forms investigated. There is intellectual
freedom to set aside and \textbf{dispose of} (apothesis) old preconceptions, prejudices and stereotypes, and also to explore and investigate new paths, to \textbf{put together} (prosthesis) and try out new concepts and ideas.

 \item \textbf{$Y_2$- Synthesis:} 
 It is at this vertex that \textbf{new laws are formulated}; this is the point of Eureka moment(s). A selection of old and and new concepts seem to click into place, fitting together in the form of new laws, laws that are able to explain new phenomena and incorporate objects of an expanded reality.

 \item \textbf{$I_2$- Enthesis:} 
 At this vertex new laws, concepts and methods must \textbf{enter and be integrated into}  a consistent and coherent system. 
 At this stage many tasks are performed in order to combine novel and traditional pieces or to accommodate original and conventional components into an well-integrated framework.
 Finally, new experimental means and methods are developed and perfected, allowing the new laws to be corroborated.

 \item \textbf{$A_2$- New Thesis:}
 At this vertex, the new theory is \textbf{accepted} as the standard paradigm that succeeds the preceding one ($A_1$). Acceptance occurs after careful determination of fundamental constants and calibration factors (including their known precision), metrological and instrumentational error bounds, etc. 
 At later stages of maturity, equivalent theoretical frameworks may be developed using alternative formalisms and ontologies.
 For example, analytical mechanics offers variational alternatives that are (almost) equivalent to the classical formulation of Newtonian mechanics \citep{Abraham1987}.  Usually, these alternative world-views reinforce the trust and confidence on the underlying laws. 
 Nevertheless, the existence of such alternative perspectives may also foster exploratory efforts and investigative works in the next cycle in evolution. 
\end{itemize}

\Cref{table:spiral} applies this spiral structure to
the evolution of the theories of
orbital astronomy and chemical affinity.
The evolution of orbital astronomy has
been widely studied \citep{Hawking2004}.
The evolution of chemical affinity is
presented in greater detail in \citet{Stern2014b, Stern2014c}.

\begin{table}
 \centering
 \begin{tabular}{lll}
  \hline 
  Vertex & Orbital astronomy & Chemical Affinity \\ 
  \hline \hline 
  $I_1$- Enthesis/  & 
  Ptolemaic/ Copernican &  
  Geoffroy affinity table and \\  
  $A_1$- Thesis & 
  cycles and epicycles & 
  highest rank substitution \\ 
  \hline 
  $U_1$- Analysis &
  Circular or oval orbits? &
  Ordinal or numeric affinity? \\ 
  \hline 
  $E_1$- Antithesis &
  Non-circular orbits &
  Non-ordinal affinity \\ 
  \hline 
  $O_2$- Apothesis &
  Elliptic planetary orbits, &
  Integer affinity values, \\
   \mbox{} /Prosthesis &
  focal centering of sun &
  for arithmetic recombination \\ 
  \hline 
  $Y_2$- Synthesis &
  Kepler laws! &
  Morveau rules and tables! \\ 
  \hline 
  $I_2$- Enthesis &
  Vortex physics theories, &
  Affinity + stoichiometry \\
  $A_2$- Thesis &
  Keplerian astronomy &
  substitution reactions \\ 
  \hline 
  $U_2$- Analysis &
  Tangential or radial forces? &
  Total or partial reaction? \\ 
  \hline 
  $E_2$- Antithesis &
  Non-tangential forces &
  Non-total substitutions \\
  \hline
  $O_3$- Apothesis &
  Radial attraction forces, &
  Reversible reactions,  \\ 
  \mbox{}  /Prosthesis &  
  inverse square of distance &  
  equilibrium conditions \\ 
  \hline 
  $Y_3$- Synthesis &  
  Newton laws! &  
  Mass-Action kinetics! \\ 
  \hline 
  $I_3$- Enthesis/ &  
  Newtonian mechanics \& &  
  Thermodynamic theories \\
  $A_3$- Thesis &
  variational equivalents &
  for reaction networks \\ 
  \hline 
 \end{tabular}  
 \caption{Evolution of orbital astronomy and
 chemical affinity.}
 \label{table:spiral}
\end{table} 

The above spiral structure highlights that
a statistical methodology should be able to
obtain each of the six modalities in
the hexagon of oppositions.
Before an acceptance vertex (A) in the 
hexagon is reached by the spiral of scientific evolution, theoretically precise or sharp hypotheses must be formulated. 
However, a logically coherent hypothesis test,
such as the GFBST, can choose solely between
rejecting or remaining agnostic (i.e. corroborating) such sharp hypotheses.
Once the evolving theory becomes (part of) a well-established paradigm, the GFBST can be used with the goal of accepting non-sharp hypotheses in the context of the same  paradigm, a context that includes fundamental constants and  calibration factors (and their respective uncertainties),  metrological error bounds, specified accuracies of scientific intrumentation, etc.
The non-sharp versions of sharp hypotheses used in such tests are called pragmatic, and their formulation is developed in the following sections. 

\section{Pragmatic hypotheses}
\label{sec:predictive}

In order to derive pragmatic hypotheses from precise ones,
it is necessary to define an idealized future experiment.
Let $\theta$ be an unknown parameter of interest which
is used to express scientific hypotheses and that
takes values in the parameter space, $\Theta$.
A scientific hypothesis takes the form
$H_0: \theta \in \Theta_0$, where
$\Theta_0 \subset \Theta$. Whenever
there is no ambiguity,
$H_0$ and $\Theta_0$ are used interchangeably.
Also, the determination of $\theta$ is
useful for predicting an idealized future experiment, $\Z$, 
which takes values in $\sZ$.
The uncertainty about $Z$ depends on $\theta$
by means of $P_{\ts}$, the probability measure over $\sZ$
when it is known that $\theta=\ts$, $\ts \in \Theta$.
 
Often, it is sufficient for an end-user to
determine a pragmatic hypothesis, that is, 
that the parameter lies in a set of plausible values,
which is larger than the null hypothesis.
This set can be chosen in such a way that
the variation over predictions about a future experiment is
mostly due to experimental conditions rather than to
the imprecision in the value of the parameter.
This section formally develops a methodology for
determining these pragmatic hypotheses.

In order to compare two parameter values,
we use a \emph{predictive dissimilarity}, $d_{\Z}$,
which is a function 
$d_{\Z}: \Theta \times \Theta \rightarrow \Re^{+}$,
such that $d_{\Z}(\t0,\ts)$ measures how much
the predictions made for $\Z$ based on
$\ts$ diverge from the ones made based on $\t0$.
We define and compare three possible choices for
such a dissimilarity.

\begin{definition}
 \label{ex:kl}
 The \emph{Kullback-Leibler predictive dissimilarity},
 $\mbox{KL}_{\Z}$ is
 \begin{align*}
  \mbox{KL}_{\Z}(\t0,\ts)
  &= \mbox{KL}(\P_{\ts},\P_{\t0})
  =\int_{\sZ}
  \log\left(\frac{d\P_{\ts}}{d\P_{\t0}}\right)d\P_{\ts},
 \end{align*}
 that is, $\mbox{KL}_{\Z}(\t0,\ts)$
 is the relative entropy between
 $\P_{\ts}$ and $\P_{\t0}$.
\end{definition}

\begin{example}[Gaussian with known variance]
 \label{ex:gaussKL}
 Let $\Z=(Z_1 \ldots, Z_d) \sim N(\theta,\Sigma_0)$ be
 a random vector with a multivariate Gaussian distribution:
 \begin{align*}
  \frac{d\P_\theta(\z)}{d\z}
  &= \|2\pi \Sigma_0\|^{-0.5}
  \exp\left(-0.5(\z-\theta)^{t}\Sigma_0^{-1}
  d(\z-\theta)\right) \\
  \mbox{KL}_{\Z}(\t0,\ts)
  &=\int_{\mathbb{R}^d}
  \log\left(\frac{d\P_\ts(\z)}{d\P_\t0(\z)}\right)d\P_\ts(\z)
  = 0.5(\t0-\ts)^{t}\Sigma_0^{-1}(\t0-\ts),
 \end{align*}
 
 When $d=1$ and $\Sigma_0 = \sigma^2_0$,
 \begin{align}
  \label{eq:kl}
  KL_{\z}(\t0,\ts)
  &= \frac{(\t0-\ts)^2}{2\sigma^2_0}
 \end{align}
\end{example}

The KL dissimilarity evaluates
the distance between the predictive probability distributions
for the future experiment under
two parameter values, $\theta_0$ and $\theta^*$.
Although the KL dissimilarity is general,
it can be challenging to interpret. In particular,
it can be hard to establish how good are 
the predictions for $\Z$ based on $\ts$ when
$\Z$ is actually generated from $\t0$ and
$KL_{\Z}(\t0,\ts) \leq \epsilon$.
A more interpretable dissimilarity is obtained by
taking $d_{\Z}(\t0,\ts)$ to measure how far are
the best predictions for $\Z$ based on $\ts$ and $\t0$.
In this case, if one makes a prediction for $\Z$
based on $\ts$, $\z_*$, and 
$\Z$ was actually generated using $\t0$, then
$d_{\Z}(\t0,\ts) \leq \epsilon$ guarantees that
$\z_*$ will be at most $\epsilon$ apart from
the best possible prediction.
Such a dissimilarity is discussed in
the following definition.

\begin{definition}[Best prediction dissimilarity - BP]
 \label{ex:d-pred}
 Let $\hZ: \Theta \rightarrow \sZ$ be such that
 $\hZ(\t0)$ is the best prediction for $\Z$
 given that $\theta=\theta_0$. For example,
 one can take
 \begin{align*}
  \hZ(\t0)
  &= \arg \min_{\z \in \sZ} \delta_{\Z,\t0}(\z),
 \end{align*}
 where $\delta_{\Z,\t0}:\sZ \rightarrow \mathbb{R}$
 is such that $\delta_{\Z,\t0}(\z)$ measures 
 how bad $\z$ predicts $\Z$ when $\theta=\t0$.
 The \emph{best prediction dissimilarity}, 
 $\mbox{BP}_{\Z}(\t0,\ts)$, measures 
 how badly $\hZ(\ts)$ predicts $\Z$ relatively to
 $\hZ(\t0)$ when $\theta = \t0$. Formally,
 \begin{align*}
  \mbox{BP}_{\Z}(\t0,\ts)
  &= g\left(\frac{\delta_{\Z,\t0}(\hZ(\ts))
  -\delta_{\Z,\t0}(\hZ(\t0))}
  {\delta_{\Z,\t0}(\hZ(\t0))}\right),
 \end{align*}
 where $g:\mathbb{R} \longrightarrow \mathbb{R}$ is
 a motononic function.
 The choice of $g$ in
 a particular setting aims at
 improving the interpretation
 of the best prediction dissimilarity criterion.
\end{definition}

\begin{example}[BP under quadratic form]
 \label{ex:bp-l2}
 Let $\sZ=\mathbb{R}^d$,
 $\mu_{\Z,\theta}=\E[\Z|\theta]$,
 $\Sigma_{\Z,\theta}=\V[\Z|\theta]$ and 
 $\S$ be a positive definite matrix.
 Define the quadratic form induced by $\S$ 
 to be $\|\z\|_{\S}^2 = \z^{T}\S\z$ and
 \begin{align*}
  \delta_{\Z,\t0}(\z)
  =\E\left[\|\Z-\z\|_{\S}^2|\theta=\t0\right]
 \end{align*}
 The optimal prediction under $\theta^*$ is 
 $\hZ(\ts)=\mu_{\Z,\ts}$.
 It follows that
 \begin{align*}
  \delta_{\Z,\t0}(\hZ(\ts))
  &=\E\left[\|\Z-\mu_{\Z,\ts}\|_{\S}^2 |\theta=\t0\right] \\
  &=\|\mu_{\Z,\t0}-\mu_{\Z,\ts}\|_{\S}^2
  +\E\left[\|\Z-\mu_{\Z,\t0}\|_{\S}^2|\theta=\t0\right]
 \end{align*}
 In particular,
 $\delta_{\Z,\t0}(\hZ(\t0))
 =\E\left[\|\Z-\mu_{\Z,\t0}\|_{\S}^2|\theta=\t0\right]$.
 Therefore,
 \begin{align}
  \label{eq:bp-l2-1}
  \mbox{BP}_{\Z}(\t0,\ts)
  &=g\left(\frac{\|\mu_{\Z,\t0}-\mu_{\Z,\ts}\|_{\S}^2}
  {\E\left[\|\Z-\mu_{\Z,\t0}\|_{\S}^2|\theta=\t0\right]}
  \right)
 \end{align}
 In this example, BP$_{\Z}$ can be put
 in the same scale as $\Z$ by taking
 $g(x) = \sqrt{x}$. 
 Also, two choices of $\S$ are of particular interest.
 When $\S = \V[\Z|\theta=\t0]^{-1}$,
 \cref{eq:bp-l2-1} simplifies to
 \begin{align}
  \label{eq:bp-l2-2}
  \mbox{BP}_{\Z}(\t0,\ts)
  &=g\left(d^{-1}\|\mu_{\Z,\t0}
  -\mu_{\Z,\ts}\|^2_{\Sigma^{-1}_{\Z,\t0}}\right)
 \end{align}
 Similarly, when $\S$ is the identity matrix,
 \cref{eq:bp-l2-1} simplifies to
 \begin{align}
  \label{eq:bp-l2-3}
  \mbox{BP}_{\Z}(\t0,\ts)
  =g\left(\frac{\|\E[\Z|\theta=\t0]-\E[\Z|\theta=\ts]\|_2^2}
  {\tr(\V[\Z|\theta=\t0])}\right)
 \end{align}
 \Cref{eq:bp-l2-3} admits an intuitive interpretation.
 The larger the value of $\tr(\V[\Z|\theta=\t0])$,
 the more $\Z$ is dispersed and
 the harder it is to predict its value.
 Also, $\|\E[\Z|\theta=\t0]-\E[\Z|\theta=\ts]\|^2_2$
 measures how far apart are the best prediction for 
 $\Z$ under $\theta=\t0$ and $\theta=\ts$.
 That is, $\text{BP}_{Z}(\t0,\ts)$ captures that,
 if one predicts $\Z$ assuming that 
 $\theta=\ts$ when actually $\theta=\t0$, then 
 the error with respect to the best prediction
 is increased as a function of
 the distance between the predictions over
 the dispersion of $\Z$.
\end{example}

\begin{example}[Gaussian with known variance]
 \label{ex:gaussBP}
 Consider again \Cref{ex:gaussKL} and let
 $\delta_{\Z,\t0}(\z)$ be such as in
 \cref{ex:bp-l2}.
 It follows from \cref{eq:bp-l2-3} that,
 when $\S$ is the identity matrix,
 \begin{align}
  \label{eq:gaussBP-1}
  \mbox{BP}_{\Z}(\t0,\ts)
  &= g\left(\frac{\|\t0-\ts\|_2^2}{\tr(\Sigma_0)}\right)
 \end{align}
 Similarly, it follows from \cref{eq:bp-l2-2} that,
 when $\S=\Sigma_0^{-1}$,
 \begin{align}
  \label{eq:gaussBP-2}
  \mbox{BP}_{\Z}(\t0,\ts)
  &= g\left(d^{-1}(\t0-\ts)^{t}\Sigma_0^{-1}(\t0-\ts)\right)
 \end{align}
 Conclude from \cref{eq:gaussBP-2} that,
 if $\S=\Sigma_0^{-1}$ and $g(x)=x$, then
 $\mbox{BP}_{\Z}(\t0,\ts) = 2d^{-1}\mbox{KL}_{\Z}(\t0,\ts)$.
 Also, when $d=1$, $\Sigma_0=\sigma_0^2$ and 
 $g(x)=\sqrt{x}$ both
 \cref{eq:gaussBP-1} and \cref{eq:gaussBP-2}
 simplify to 
 \begin{align}
  \label{eq:gaussBP-3}
  \mbox{BP}_{\Z}(\t0,\ts)
  &=\sigma_0^{-1}|\t0-\ts|
 \end{align}
 In some situations, $\Z$ is 
 the average of $m$ independent observations
 distributed as $N(\theta,\Sigma_0)$.
 In this case, $\Z \sim N(\theta,m^{-1}\Sigma_0)$.
 It follows from \cref{eq:gaussBP-1} that
 $BP_{\Z}(\t0,\ts) 
 = g\left(\frac{m\|\t0-\ts\|_2^2}{\tr(\Sigma_0)}\right)$,
 when $\S$ is the identity, and
 $\mbox{BP}_{\Z}(\t0,\ts)
 = g\left(md^{-1}(\t0-\ts)^{t}
 \Sigma_0^{-1}(\t0-\ts)\right)$,
 when $\S=\Sigma_{0}^{-1}$.
\end{example}

Although $\mbox{BP}_{\Z}$ is more interpretable then
$\mbox{KL}_{\Z}$ it also relies on
more tuning variables, such as
$\delta$, $\hat{Z}$ and $g$. A balance between
these features is obtained by
a third predictive dissimilarity,
which evaluates how easy it is to recover
the value of $\theta$ between
$\theta_0$ or $\theta^*$ based on $\Z$.

\begin{definition}[Classification distance - CD]
 Let $\hT_{\t0,\ts}: \sZ \rightarrow \Theta$
 be such that
 \begin{align*}
  \hT_{\t0,\ts}(\z) &=
  \arg\max_{\theta \in \{\t0,\ts\}}
  f_{\Z}(\z|\theta)
 \end{align*}
 $\hT_{\t0,\ts}$ assigns to each possible outcome of
 the future experiment $\z$, which value of $\theta$,
 $\t0$ or $\ts$, makes the experimental result more likely. 
 The \emph{classification distance} between
 $\t0$ and $\ts$, $\text{CD}(\t0,\ts)$, is defined as
 \begin{align*}
  \text{CD}(\t0,\ts)
  &= 0.5\P\left(\hT_{\t0,\ts}(\Z) = \t0|\t0\right)
  + 0.5\P\left(\hT_{\t0,\ts}(\Z) = \ts|\ts\right)
  - 0.5
 \end{align*}
 $\text{CD}(\t0,\ts)+0.5$ is the best Bayes utility in 
 an hypothesis test of $\t0$ against $\ts$
 using a uniform prior for $\theta$ and the 0/1 utility \citep{Berger2013}.
 By subtracting $0.5$ from this quantity,
 $\text{CD}(\t0,\ts)$ varies between 
 $0$ and $0.5$ and is a distance. Also,
 \begin{align*}
  \text{CD}(\t0,\ts)
  &= 0.5 \text{TV}(\P_{\t0},\P_{\ts})
  = 0.25 \|\P_{\t0}-\P_{\ts}\|_1,
 \end{align*}
 where 
 $\text{TV}(\P_{\t0},\P_{\ts})
  = \sup_{A} |\P_{\t0}(A)-\P_{\ts}(A)|$ and
 $ \|\P_{\t0}-\P_{\ts}\|_1 =
 \int_{\sZ}{|\P_{\t0}(z)-\P_{\ts}(z)|dz}$ is
 the $\mathcal{L}_1$-distance between
 probability measures.
\end{definition}

\begin{example}[Gaussian with known variance]
 \label{ex:gaussTV}
 Consider \cref{ex:gaussKL,ex:gaussBP}.
 When $d=1$, $\Sigma_0 = \sigma_0^2$, obtain
 \begin{align}
  \label{eq:CD}
  \mbox{CD}_{\Z}(\t0,\ts)
  &= \Phi \left(\frac{|\tz-\ts|}{2\sigma_0}\right)-\frac{1}{2}
 \end{align}
 Note that, in this case, $CD$ would be the same as
 $BP$ if, instead of taking $g(x) = \sqrt{x}$,
 one chose $g(x) = \Phi(0.5\sqrt{x}) - 0.5$. 
\end{example}

Although analytical expressions for 
\text{CD} are generally not available,
it is possible to approximate
it via numerical integration methods.

\subsection{Singleton hypotheses}
\label{sec:simple}

We start by defining the pragmatic hypothesis
associated to a singleton hypothesis.
A singleton hypothesis is one in which the
parameter assumes a single value,
such as $H_0:\theta=\theta_0$.
In this case, the pragmatic hypothesis 
associated to $H_0$ is the set of points whose
dissimilarity to $\theta_0$ is at most $\epsilon$,
as formalized below.

\begin{definition}[Pragmatic hypothesis for a singleton]
\label{def:pragSimple}
 Let $H_{0}:\theta=\t0$,
 $d_{\Z}$ be a predictive dissimilarity function
 and $\epsilon>0$.
 The pragmatic hypothesis for $H_0$,
 $Pg(\{\t0\},d_{\Z},\epsilon)$, is
 \begin{align*}
  Pg(\{\t0\},d_{\Z},\epsilon)
  &= \{\ts \in \Theta:d_{\Z}(\t0,\ts) \leq \epsilon\}
 \end{align*}
\end{definition}

\begin{example}[Gaussian with known variance]
 \label{ex:gauss}
 Consider \cref{ex:gaussKL,ex:gaussBP} when
 $d=1$, $\Sigma_0 = \sigma_0^2$ and 
 $g(x)=\sqrt{x}$. It follows from
 \cref{eq:kl,eq:gaussBP-3,eq:CD} that
 \begin{align*}
  Pg(\{\t0\},BP_{\Z},\epsilon)
  &= \left[\t0 -\epsilon\sigma_0,
  \t0 +\epsilon\sigma_0\right] \\
  Pg(\{\t0\},KL_{\Z},\epsilon) 
  &= \left[\t0 -\sqrt{2\epsilon}\sigma_0,
  \t0 +\sqrt{2\epsilon}\sigma_0\right] \\ 
  Pg(\{\t0\},CD_{\Z},\epsilon)
  &= \left[\t0 -2\Phi^{-1}(0.5+\epsilon)\sigma_0,
  \t0 + 2\Phi^{-1}(0.5+\epsilon)\sigma_0\right]
 \end{align*}
 
 Note that the size of each of the pragmatic hypothesis is
 proportional to $\sigma_0$. This occurs because
 every one of the predictive dissimilarity functions makes
 the prediction error due to the unknown parameter value small
 with respect to that due to the data variability, $\sigma^2_0$.
\end{example}

\subsection{Composite hypotheses}
\label{sec:composite}

Next, we consider pragmatic hypotheses for
general hypotheses $H_0: \theta \in \Theta_0$,
where $\Theta_0 \subset \Theta$.

\begin{definition}
 For each hypothesis $\Theta_0 \subseteq \Theta$,
 predictive dissimilarity $d_{\Z}$ and
 $\epsilon > 0$, $Pg(\Theta_0, d_{\Z}, \epsilon)$ is
 the pragmatic hypothesis associated to $\Theta_0$
 induced by $d_{\Z}$ and $\epsilon$.
 Whenever $d_{\Z}$ and $\epsilon$ are clear or
 not relevant to the result, we write $Pg(\Theta_0)$ instead of
 $Pg(\Theta_0, d_{\Z}, \epsilon)$.
\end{definition}

\begin{figure}
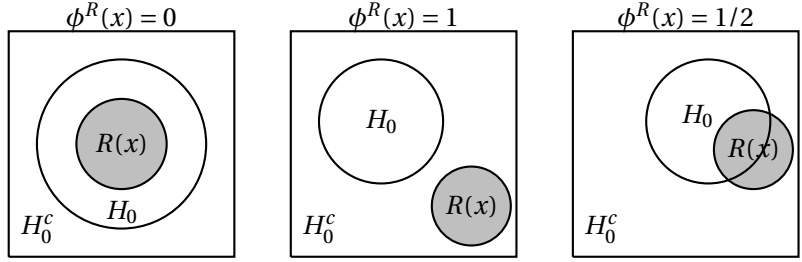

 \center
 \gfbstfig
 \mbox{} \vspace{-2mm} \mbox{} \\ 
 \caption{$\phi(x)$ is
 an agnostic test based on 
 the region estimator $R(x)$ 
 for testing $H_0$.}
 \label{fig::region}
\end{figure}

In order to construct these pragmatic hypotheses,
we use logically coherent agnostic hypothesis tests.
For each hypothesis, an agnostic hypothesis test can
either reject it (1), accept it (0) or remain agnostic (1/2).
\citet{Esteves2016} shows that 
an agnostic hypothesis test is logically coherent
if and only if it is based on a region estimator.
Such tests are presented in
\cref{def::region} and illustrated in \cref{fig::region}.

\begin{definition}
 Let $\sX$ denote the sample
 space of the data used to
 test a hypothesis.
 A region estimator is a function,
 $R: \sX \longrightarrow \mathcal{P}(\Theta)$,
 where $\mathcal{P}(\Theta)$
 is the power set of $\Theta$.
\end{definition}

\begin{definition}[Agnostic test based on a region estimator]
 \label{def::region}
 The agnostic test based on the region estimator $R$ for
 testing $H_0$, $\phi_{H_0}^R$, is such that
 \begin{align*}
  \phi_{H_0}^R(x) &=
  \begin{cases}
   0 & \text{, if } R(x) \subseteq H_0 \\
   1 & \text{, if } R(x) \subseteq H_0^c \\
   \half & \text{, otherwise.} \\
  \end{cases}
 \end{align*}
\end{definition}

Besides the logical conditions on
the hypothesis test, one might also impose
 logical restraints on 
how pragmatic hypotheses are constructed.
For instance, let $A$ and $B$ be two hypothesis such that
$B$ logically entails $A$, that is, $B \subseteq A$.
If a logically coherent test accepts $B$, then
it also accepts $A$. This property is 
called monotonocity \citep{Izbicki2015,Silva2015,Fossaluza2017}.
One might also impose that $Pg$ is such that,
if a logically coherent hypothesis test
accepts $Pg(B)$, then it should also accept $Pg(A)$.
Similarly, let $(A_i)_{i \in I}$ be a collection
of hypothesis which cover $A$, that is,
$A \subseteq \cup_{i \in I}A_i$.
If a logically coherent hypothesis test rejects
every $A_i$, then it rejects $A$.
This property is called union consonance.
One might also impose that $Pg$ is such that,
if a logically coherent hypothesis test 
rejects $Pg(A_i)$ for every $i$, then
it should also reject $Pg(A)$.
The above conditions define
the logical coherence of a procedure for
constructing pragmatic hypotheses.

\begin{definition}
 \label{defn:logical-pragmatics}
 A procedure for constructing pragmatic hypothesis, $Pg$,
 is logically coherent if, for every
 logically coherent hypothesis test $\phi$
 and sample point $x$:
 \begin{enumerate}
  \item If $\phi_{Pg(B)}(x) = 0$
  for some $B \subseteq A$, then
  $\phi_{Pg(A)}(x) = 0$.
  
  \item If $\phi_{Pg(A_i)}(x) = 1$
  for every $i \in I$ and
  $A \subseteq \cup_{i \in I}A_i$, then
  $\phi_{Pg(A)}(x) = 1$.
 \end{enumerate}
\end{definition}

In order to motivate the above definition,
consider that the frequencies of $AA$, $AB$ and $BB$
in a given population are $\theta_1$, $\theta_2$ and $\theta_3$.
Note that $B := \{0.25,0.5,0.25\}$ is a subset of
$A = \{(p^2, 2p(1-p),(1-p)^2): p \in [0,1]\}$, which denotes
the Hardy-Weinberg equilibrium. That is,
if the frequencies $AA$, $AB$ and $BB$ are,
respectively, 0.25, 0.5 and 0.25, then
the population follows the Hardy-Weinberg equilibrium.
As a result, if one pragmatically accepts that
the population satisfies the specified proportions, then
one might also wish to pragmatically accept that
the population follows the Hardy-Weinberg.
Similarly, if one pragmatically rejects 
for every $p \in [0,1]$ that
the frequencies of $AA$, $AB$ and $BB$ are,
respectively, $p^2$, $2p(1-p)$ and $(1-p)^2$, then
one might also wish to pragmatically reject that
the population follows the Hardy-Weinberg equilibrium.
These conditions are assured in 
\cref{defn:logical-pragmatics}.

In a logically coherent procedure for
constructing pragmatic hypotheses,
the pragmatic hypothesis associated
to a composite hypothesis is
completely determined by the pragmatic hypotheses
associated to simple hypotheses.
This result is presented in \cref{thm:union}.

\begin{theorem}
 \label{thm:union}
 A procedure for constructing pragmatic hypothesis, $Pg$,
 is logically coherent if and only if,
 for every hypothesis $\Theta_0$,
 $Pg(\Theta_0) = \bigcup_{\theta \in \Theta_0}Pg(\{\theta\})$.
\end{theorem}

Using \Cref{thm:union} it is possible to determine
a logically coherent procedure for
constructing pragmatic hypotheses by
determining only the pragmatic hypothesis
associated to simple hypothesis,
such as in \cref{sec:simple}.
\Cref{thm:union} is illustrated in \cref{sec:applications}.

Besides being logically coherent,
it is often desirable in statistics 
\citep{Pereira2008a,Stern2014} and
in science \citep{Stern2011b,Stern2017a} for
a procedure to be invariant to reparametrization.
That is, that the procedure reaches the same conclusions
whatever the coordinate system  
is used to specify both the sample and the parameter spaces.
For instance, the pragmatic hypothesis that is obtained
using the International metric system should be
compatible to the one that is obtained using
the English metric system.
Invariance to reparametrization is
formally presented in \cref{defn:invariance}.

\begin{definition}
 $\left(\P^*_{\theta^*}\right)_{\theta^* \in \Theta^*}$ 
 is a reparameterization of 
 $\left(\P_{\theta}\right)_{\theta \in \Theta}$
 if there exists a bijective function,
 $f: \Theta \rightarrow \Theta^*$,
 such that for every $\theta \in \Theta$,
 $\P_{\theta} = \P^*_{f(\theta)}$.
\end{definition}

\begin{definition}
 \label{defn:invariance}
 Let $\left(\P^*_{\theta^*}\right)_{\theta^* \in \Theta^*}$
 be a reparametrization of
 $\left(\P_{\theta}\right)_{\theta \in \Theta}$ by
 a bijective function, $f:\Theta \rightarrow \Theta^*$.
 Also, let $d_{\Z}$ and $d^*_{\Z}$ be
 predictive dissimilarity functions.
 The functions $d_{\Z}$ and $d^*_{\Z}$ are
 invariant to the reparametrization if
 for every logically coherent procedure for
 constructing pragmatic hypotheses, $Pg$,
 \begin{align*}
  f[Pg(\Theta_0, d_\Z, \epsilon)]
  &= Pg(f[\Theta_0], d_\Z^*, \epsilon),
 \end{align*}
\end{definition}

\Cref{defn:invariance} states that,
if $\Theta_0$ is an hypothesis and
invariance to reparametrization holds, then
the pragmatic hypothesis obtained in
a reparametrization of $\Theta_0$, say $Pg(f[\Theta_0])$, is
the same as  the transformed
pragmatic hypothesis associated to $\Theta_0$,
$f[Pg(\Theta_0)]$.
\Cref{thm:invariance} presents a
sufficient condition for obtaining
invariance to reparametrization.

\begin{theorem}
 \label{thm:invariance}
 Let $\left(\P^*_{\theta^*}\right)_{\theta^* \in \Theta^*}$
 be a reparameterization of 
 $\left(\P_{\theta}\right)_{\theta \in \Theta}$
 given by a bijective function, $f$. 
 If $d_\Z$ and $d_\Z^*$ satisfy 
 $d_\Z(\t0,\theta) = d_\Z^*(f(\t0),f(\theta))$,
 then $d_\Z$ and $d_\Z^*$ are
 invariant to this reparametrization.
\end{theorem}

\begin{corollary}
 If $d_\Z$ and $d_\Z^*$ are
 the same choice between KL, BP or CD, then
 $d_\Z$ and $d_\Z^*$ are invariant to
 every reparametrization.
\end{corollary}

The procedures for constructing pragmatic 
hypotheses
induced by $KL$ and $CD$ also
satisfy an additional property given by
\cref{thm:convergence}.

\begin{theorem}
 \label{thm:convergence}
 Let \ $\Z_m=(Z_1,\ldots,Z_m)$,
 where $Z_i$'s are i.i.d.
 $F_\theta$ and 
 $\left(F_\theta\right)_{\theta \in \Theta}$ is
 identifiable \citep{casella2002statistical,Wechsler2013}.
 Also, let $KL_m$ and $CD_m$ be
 the dissimilarities calculated using $\Z_m$.
 If $Pg$ is logically coherent then,
 for every $\Theta_0 \subseteq \Theta$ and $\epsilon > 0$,
 \begin{enumerate}[label=(\roman*)]
 \item $\left(Pg(\Theta_0,KL_m,\epsilon)\right)_{m \geq 1}$ and
 $\left(Pg(\Theta_0,CD_m,\epsilon)\right)_{m\geq 1}$ are
 non-increasing sequences of sets
 \item $Pg(\Theta_0,KL_m,\epsilon)
 \xrightarrow{m \rightarrow \infty} \Theta_0$ and
 $Pg(\Theta_0,CD_m,\epsilon)
 \xrightarrow{m \rightarrow \infty} \Theta_0$.
\end{enumerate}
\end{theorem}

\Cref{thm:convergence} states that
the sequence of pragmatic hypotheses for
$\Theta_0$ induced by $d_{\Z_m}$
is non-increasing if the dissimilarity is
evaluated by either KL or CD. 
The greater the number of observable quantities $\Z_m$,
the easier it is to distinguish
two parameter values
$\theta_0$ and $\theta^*$ and, therefore,
the smaller the amount of parameters that
are taken as close to $\t0$.
Also, as the sample size goes to infinity,
the pragmatic hypothesis associated to $\Theta_0$
converges to to $\Theta_0$.
In other words, for each $\theta_0 \in \Theta_0$,
 no other parameter value can 
predict infinitely many observable quantities with
a precision sufficiently close to that of
 $\t0$.

\section{Applications}
\label{sec:applications}

In the following,
pragmatic hypotheses for
standard statistical problems are
derived.

\begin{example}[Gaussian with unknown variance]
 \label{ex:gaussUnknown}
 Consider the setting from \Cref{ex:gauss},  
 but with $\sigma^2$ unknown and $0 < \sigma^2 \leq M^2$.
 In this case, the parameter is
 $\theta = (\mu,\sigma^2)$.
 Consider the composite hypothesis
 $H_0: \{\mu_0\} \times (0,M^2]$,
 which is often written as $H_0: \mu = \mu_0$.
 In this case let
 $\t0 = (\mu_0,\sigma_0^2)$ and
 $\Theta_0 = \{\mu_0\} \times (0,M^2]$.
 Proceeding as in \Cref{ex:gauss}, it follows that
 \begin{align*}
  Pg(\{\t0\},BP_{\Z},\epsilon)
  &= [\mu_0 - \epsilon \sigma_0, \mu_0 + \epsilon \sigma_0]
  \times (0, M^2]  \\ 
  Pg(\Theta_0, BP_{\Z}, \epsilon)
  &= [\mu_0 - \epsilon M, \mu_0 + \epsilon M]
  \times (0, M^2]&\text{\Cref{thm:union}}
 \end{align*}
 The rectangular shape of these pragmatic hypotheses seems to
 be unreasonable as, for instance, whether a
 point $(\mu,\sigma^2)$ is close to 
 $(\mu_0, \sigma_0^2)$ does not depend on
 $\sigma_0^2$. This is a consequence of
 the choice of $\delta$ in \Cref{ex:gauss}.
 
 \Cref{fig::normKL} presents
 the pragmatic hypotheses for
 $H_0: \mu = 0, \sigma^2 = 1$ and
 $H_0: \mu = 0$ when
 $\epsilon = 0.1$ and $M^2=2$, and
 using the KL and CD dissimilarities.
 Contrary to BP, the hypotheses obtained from
 these dissimilarities do not have
 a rectangular shape. In particular,
 the triangular shape of
 the pragmatic hypotheses for $H_0: \mu = 0$
 is such that the closer $\sigma^2$ is to $0$,
 the smaller the range of values for $\mu$ that
 are included in the pragmatic hypothesis.
 This behavior might be desirable since,
 when $\sigma^2$ is small, there is little
 uncertainty about the value of $\Z$ and, consequently, a narrow interval of values of $\mu$
 can predict $Z$ with precision $\epsilon$.
 
 \begin{figure}
  \centering
  \includegraphics[width=0.5\linewidth]{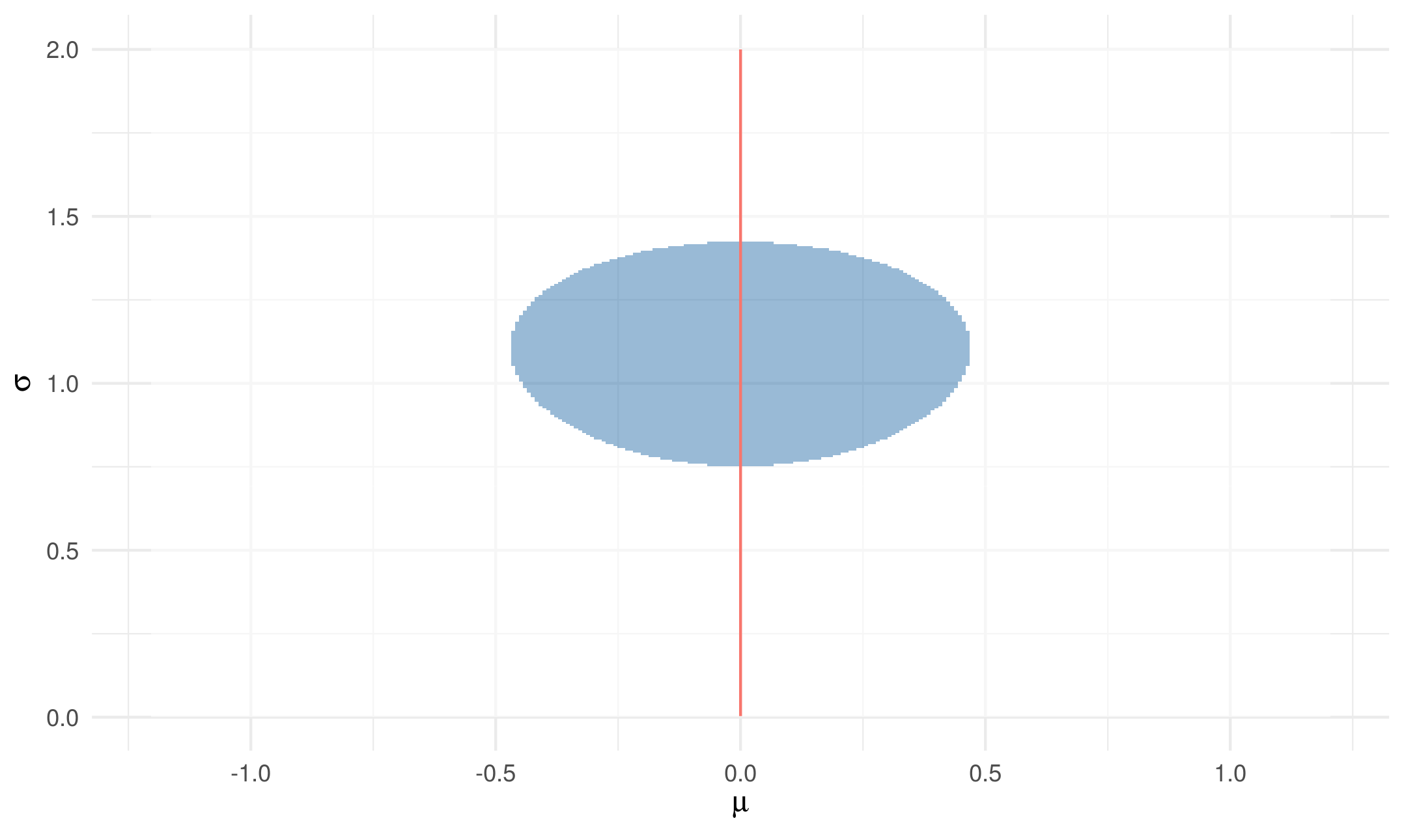}%
  \includegraphics[width=0.5\linewidth]{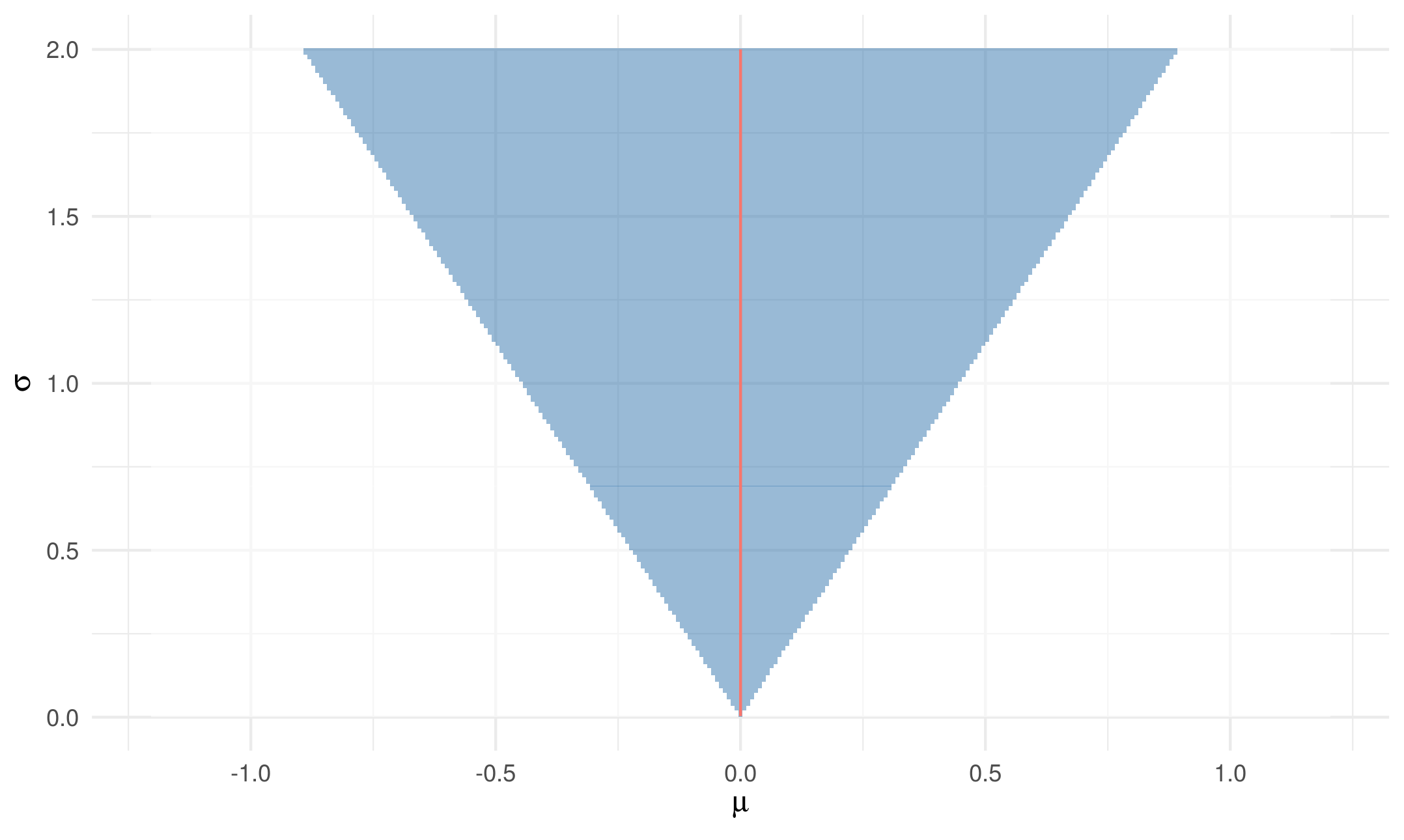}
  \includegraphics[width=0.5\linewidth]{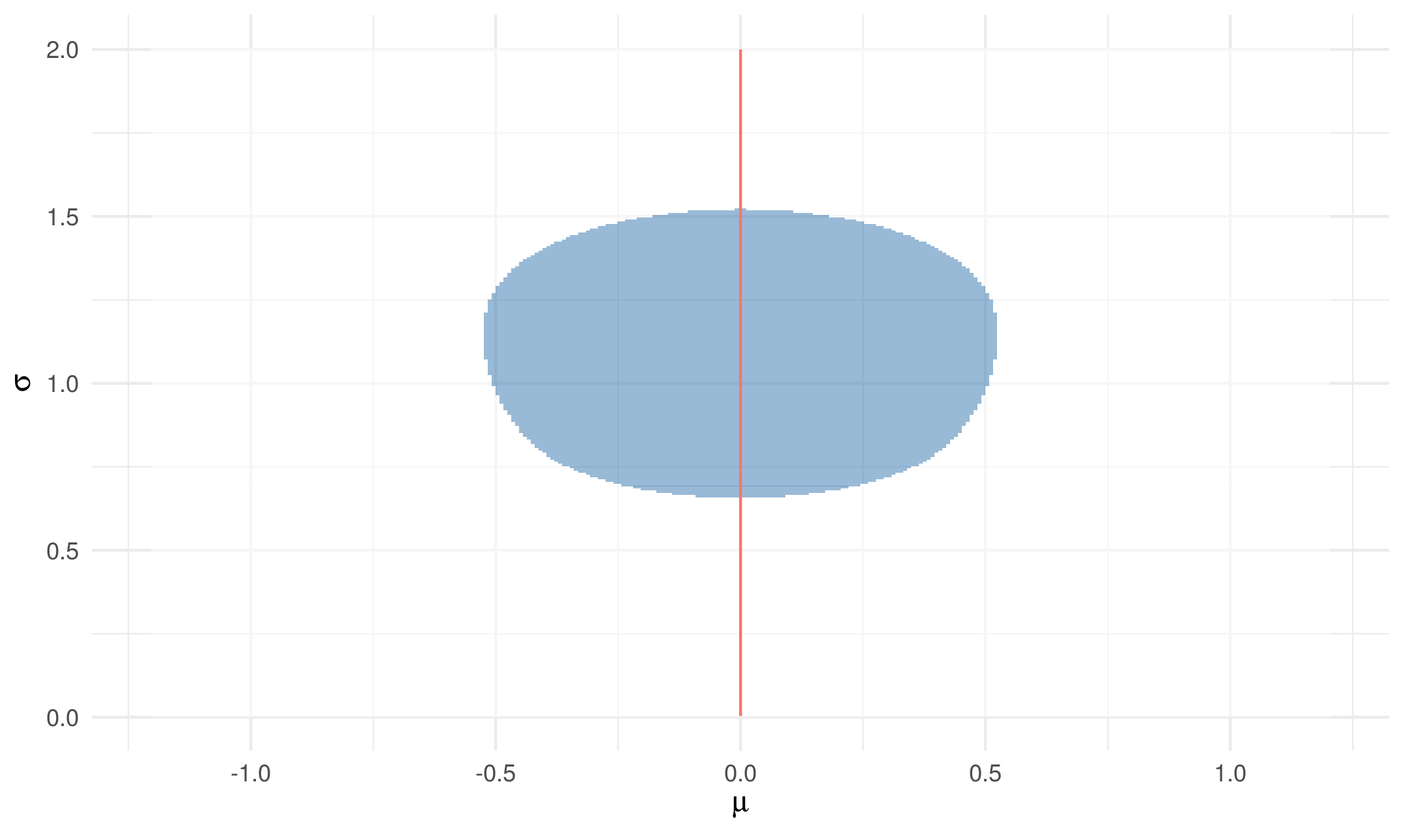}%
  \includegraphics[width=0.5\linewidth]{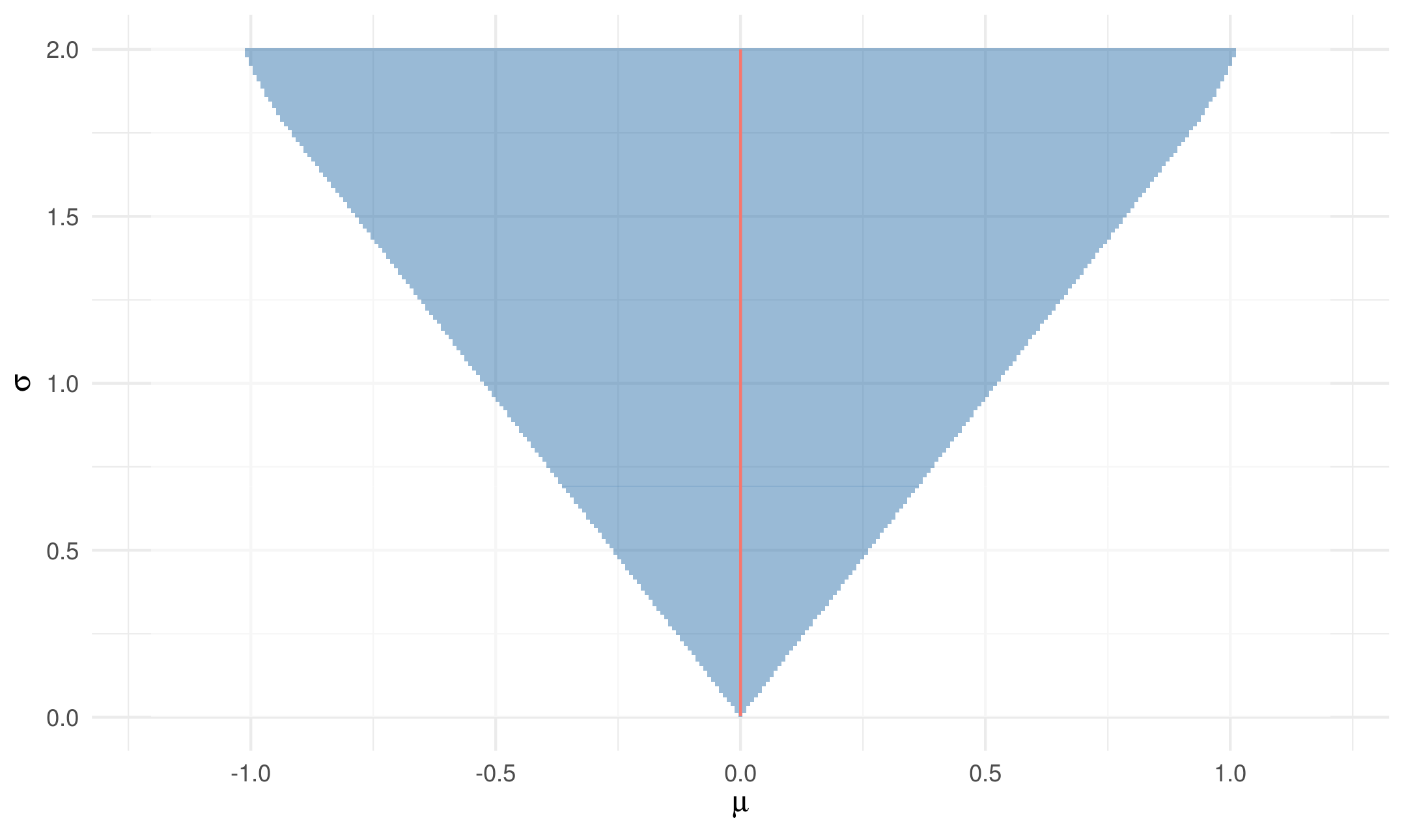}%
  \caption{Pragmatic hypotheses in
  \cref{ex:gaussUnknown} for $H_0: \mu = 0$
  with KL (upper), CD (lower), 
  $\epsilon = 0.1$, and $M^2 = 2$.
  $H_0$ is represented by a red line in all figures.
 }
  \label{fig::normKL}
 \end{figure}
\end{example}

\begin{example}[Hardy-Weinberg equilibrium] 
 Let $\Z \sim \mbox{Multinomial}(m,\btheta)$, 
 where $\btheta=(\theta_1,\theta_2,\theta_3)$, $\theta_i \geq 0$, and
 $\sum_{i=1}^3\theta_i=1$. 
 The Hardy-Weinberg (HW) hypothesis \citep{Hardy2003},
 $H_0$, which is depicted in the red curve in
 \cref{fig::hw} satisfies
 \begin{align*}
  H_0: \theta \in \Theta_0 ,&&
  \Theta_0 = \left\{\left(p^2, 2p(1-p), (1-p)^2\right)
  : 0 \leq p \leq 1\right\}
 \end{align*}
 If $\btheta_0^p = (p^2,2p(1-p),(1-p)^2)$,
 $\delta_{\Z}(\z)=\E[\|\Z-\z\|_2^2|\btheta=\btheta_0^p]$
 and $g(x) = \sqrt{x}$, then
 it follows from \cref{ex:bp-l2} that
 \begin{align*}
  \mbox{BP}_{\Z}(\btheta_0^p,\btheta^*)
  &= \left( m\times \frac{(\theta_1-p^2)^2+
  (\theta_2-2p(1-p))^2+(\theta_3-(1-p)^2)^2}
  {p^2(1-p^2)+2p(1-p)(1-2p(1-p))+(1-p)^2(1-(1-p)^2)}\right)^{0.5}
 \end{align*}
 The pragmatic hypotheses that are obtained
 using $KL$, $BP$ and $CD$ for the HW hypothesis are
 depicted in \cref{fig::hw}.
 The choice between BP or KL and CD has a large impact
 over the shape of the pragmatic hypotheses.
 While for BP the width of the pragmatic hypothesis is
 approximately uniform along the HW curve,
 the width of the pragmatic hypotheses obtained
 using $KL$ and $CD$ is smaller towards the
 edges of the HW curve. This behavior could be
 expected since, towards the edges of the HW curve,
 $\Z$ has the smallest variability.
 The figure also depicts the challenge in
 calibrating $KL$. While the pragmatic hypotheses for
 $BP$ and $CD$ have similar sizes when using 
 $\epsilon = 0.1$, this result was obtained for
 $KL$ while using $\epsilon = 0.01$. 
 \begin{figure}
  \centering
  {\includegraphics[width=0.33\linewidth]
  {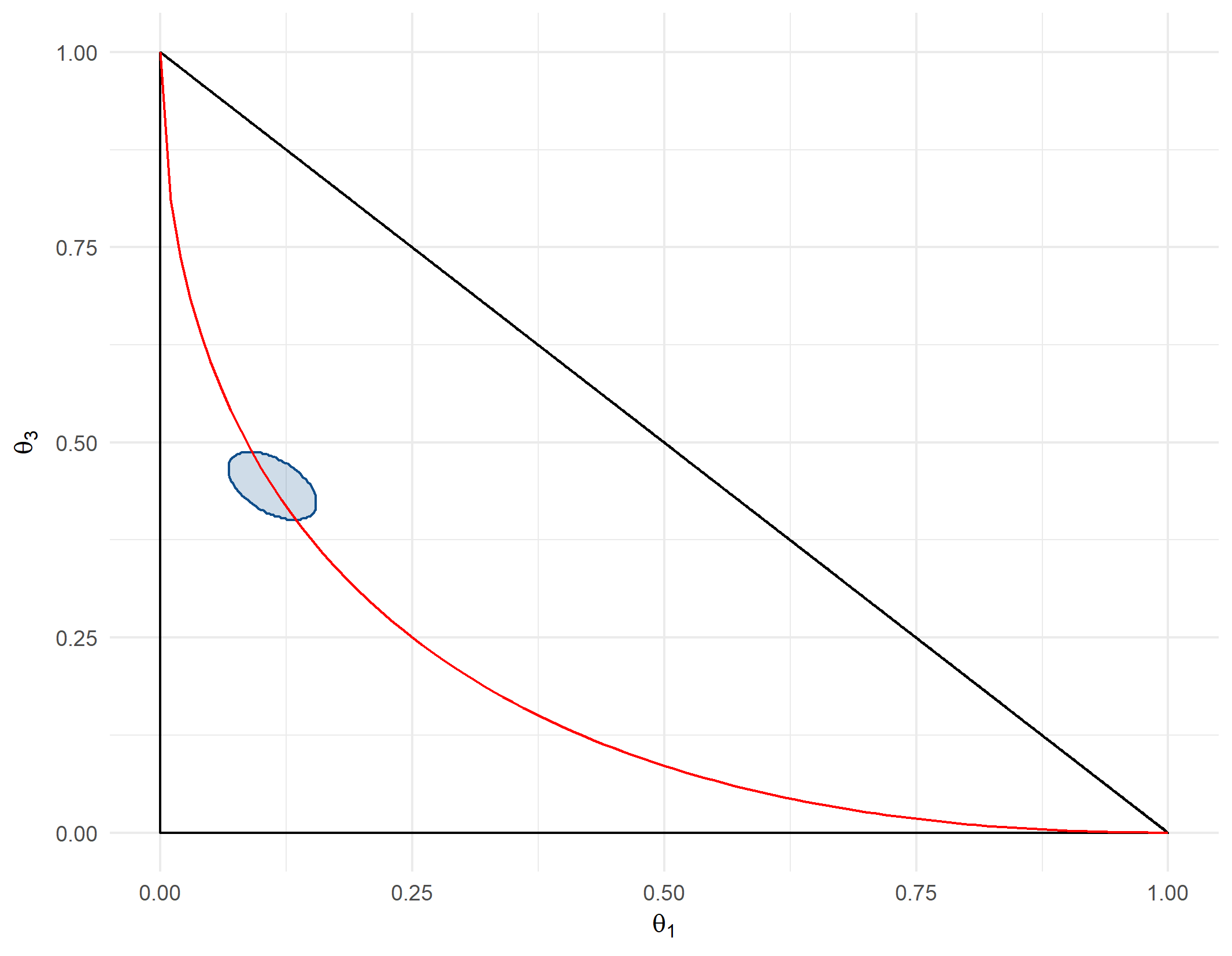}}%
  {\includegraphics[width=0.33\linewidth]
  {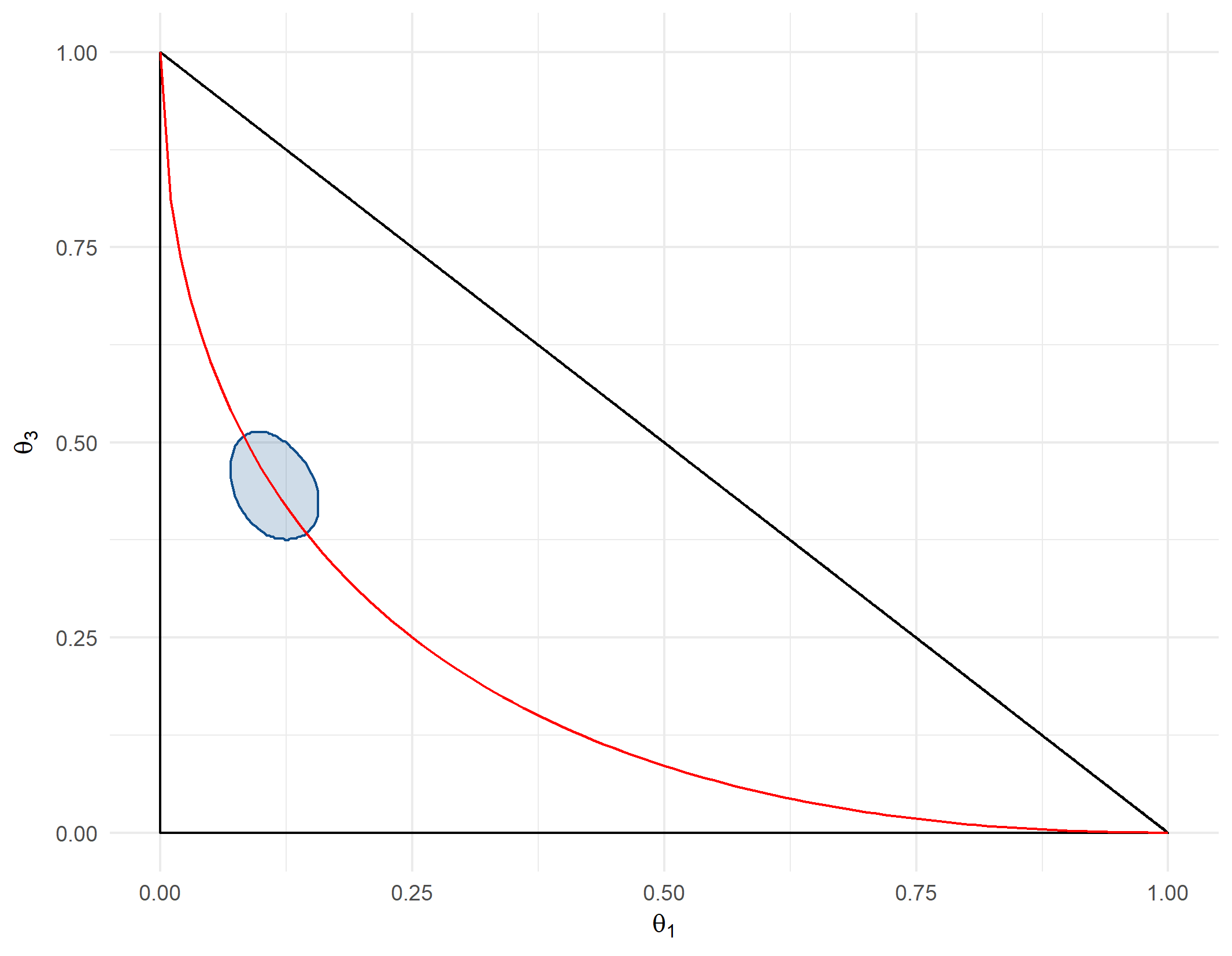}}%
  {\includegraphics[width=0.33\linewidth]
  {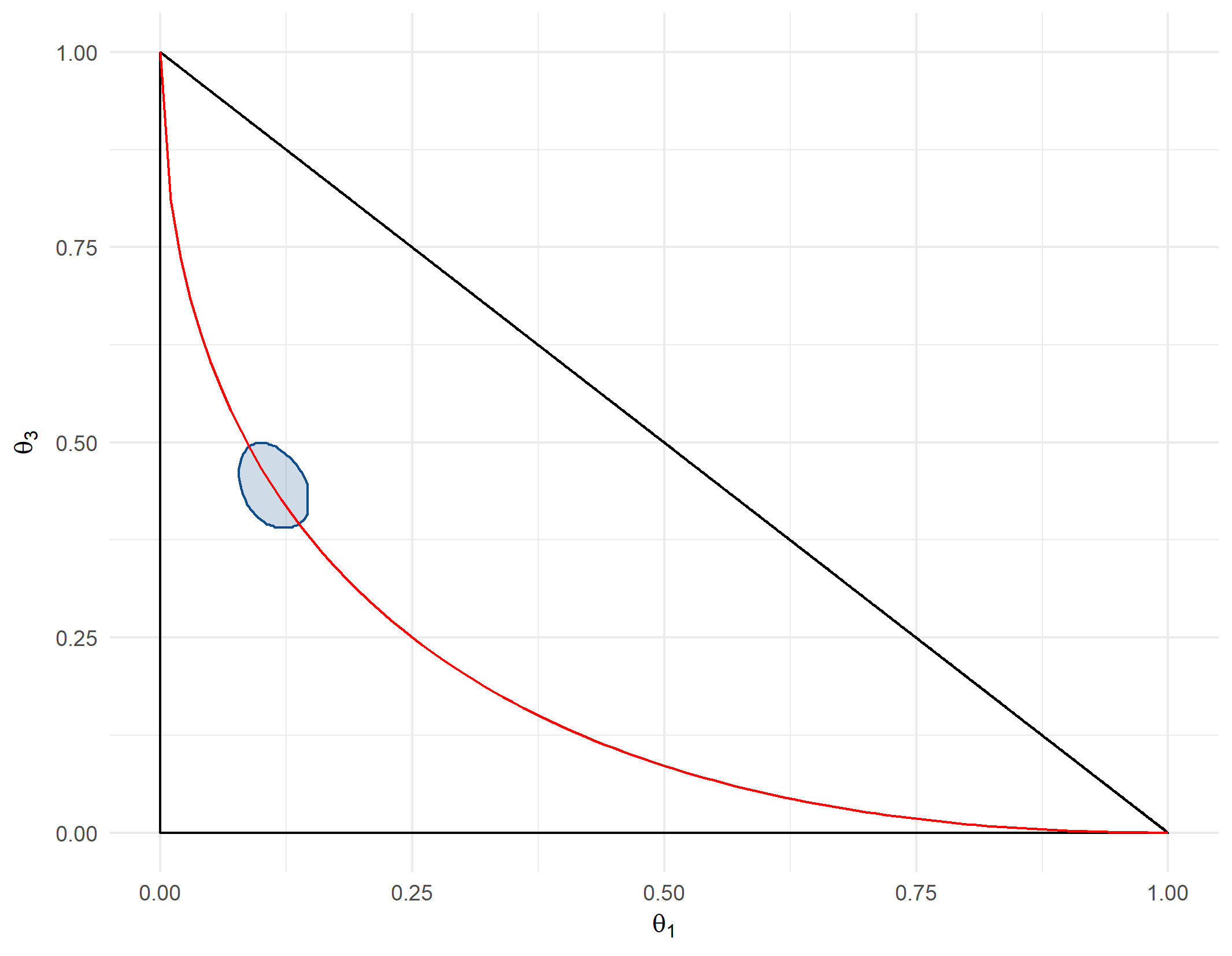}} \\
  {\includegraphics[width=0.33\linewidth]
  {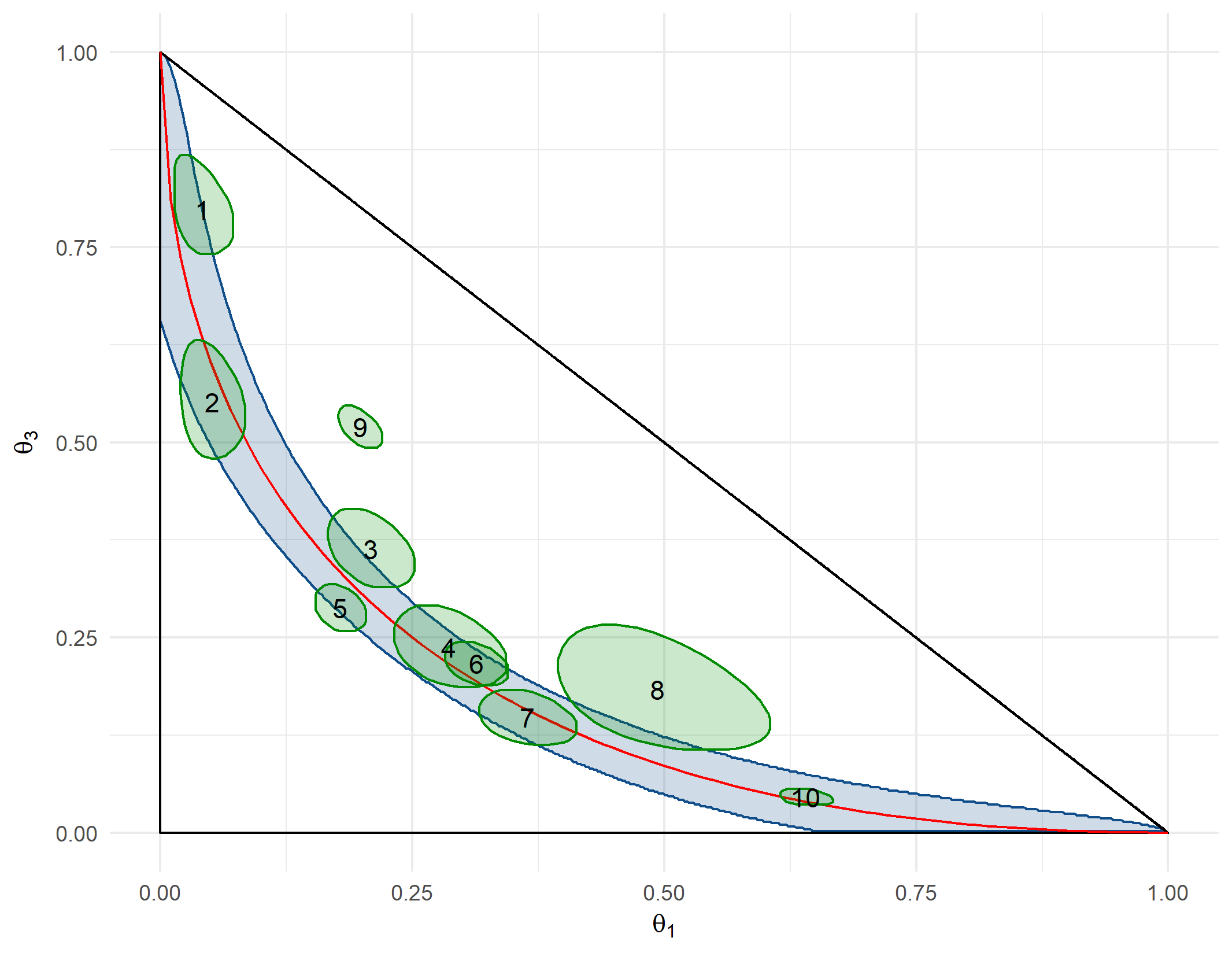}}%
  {\includegraphics[width=0.33\linewidth]
  {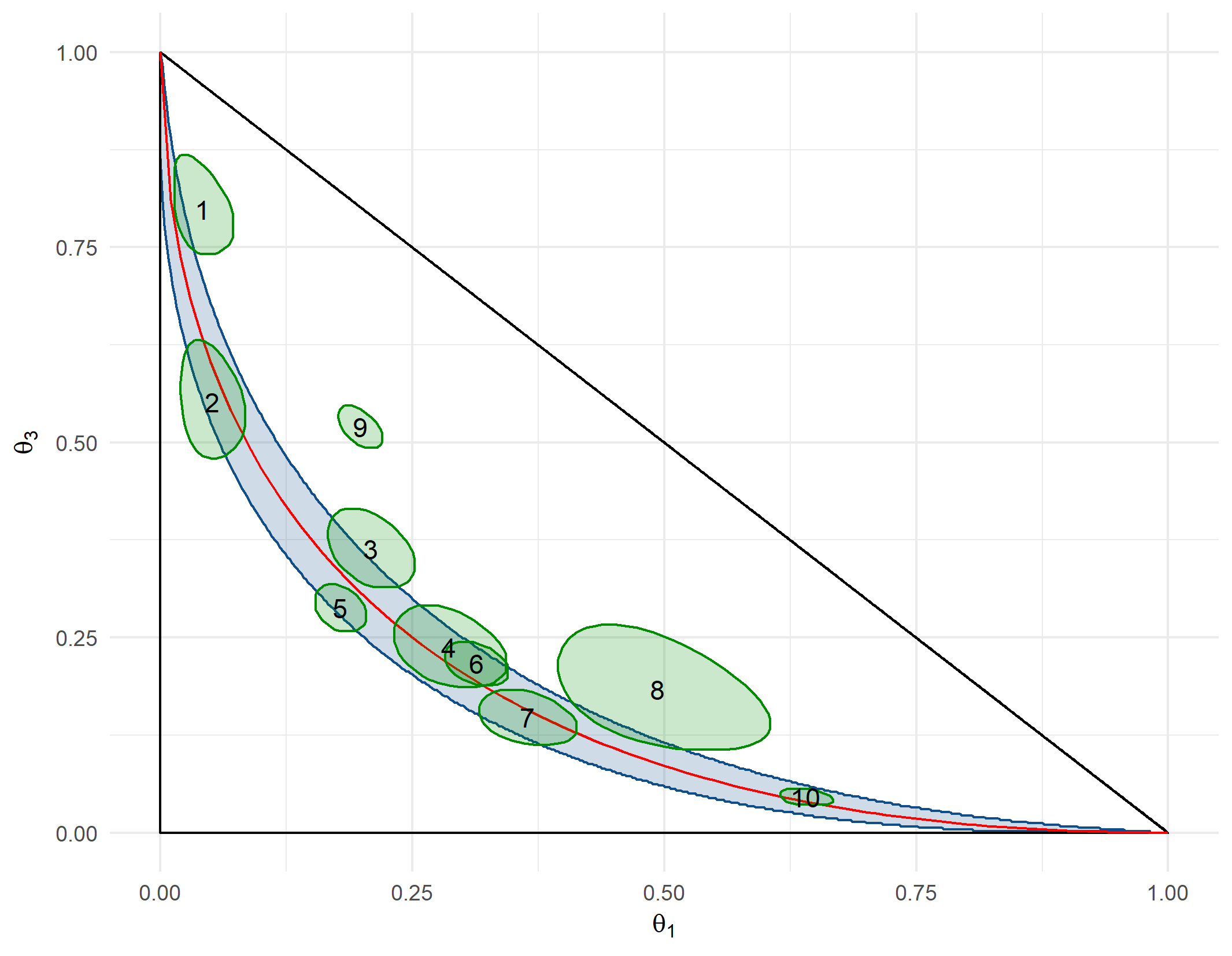}}%
  {\includegraphics[width=0.33\linewidth]
  {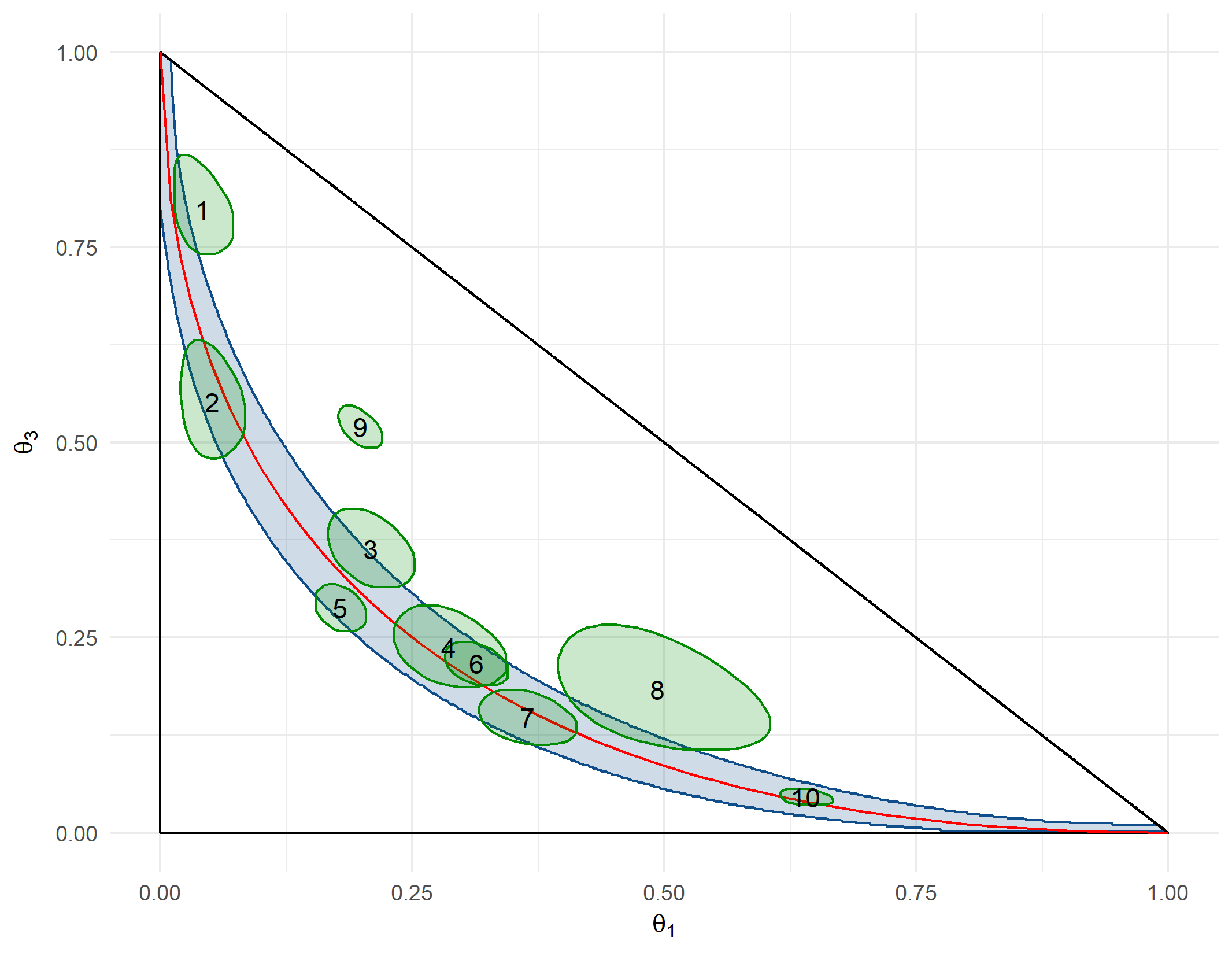}} \\ 
  \caption{Pragmatic hypotheses obtained for
  the HW equilibrium, depicted in red,
  using $m=20$, $\epsilon=0.1$ for BP and CD and
  $\epsilon=0.01$ for KL.
  The blue regions indicate the pragmatic hypothesis
  for HW and $p=\frac{1}{3}$ (top) and for HW (bottom).
  The lower, middle and right panels were obtained,
  respectively, with BP, KL and CD.
  The green regions in the right panels represents
  80\% HPD regions for the genotype distribution
  of each of the eight groups collected
  by \citet{Brentani2011} and
  two simulated datasets.}
  \label{fig::hw}
 \end{figure}
 The pragmatic hypotheses in \cref{fig::hw} are
 further tested using data from
 \citet{Brentani2011}, which is presented in \cref{tab:hw}.
 This study had the goal of 
 verifying association between 
 the APOE-$\epsilon$4 gene and Alzheimer disease.
 The lower panels of \Cref{fig::hw} present 
 the 80\% HPD regions for the distribution of 
 this gene in each of the eight groups 
 observed in the study. 
 Additionally, they present two simulated datasets, 
 9 and 10. Groups $9$ and $10$ were generated by
 populations that were, respectively,
 not under and under the HW equilibrium.
 Group $9$ and $10$ fall, respectively,
 outside and inside of the pragmatic hypothesis.
 \begin{table}
  \centering
  \begin{tabular}{rrrrl}
   \hline
   & AA & AD & DD & Decision \\ 
   \hline
   1 & 4 & 18 & 94 & Agnostic \\ 
   2 & 6 & 53 & 74 & Accept \\ 
   3 & 57 & 118 & 100 & Agnostic \\ 
   4 & 58 & 97 & 48 & Agnostic \\ 
   5 & 120 & 361 & 194 & Agnostic \\ 
   6 & 206 & 309 & 142 & Accept \\ 
   7 & 110 & 148 & 44 & Accept \\ 
   8 & 34 & 22 & 12 & Agnostic \\ 
   9 & 198 & 282 & 520 & Reject \\ 
   10 & 641 & 314 & 45 & Accept \\ 
   \hline
  \end{tabular}
  \caption{Genotype counts for the eight groups in 
  \citet{Brentani2011}. Also, the decision of 
  the GFBST agnostic hypothesis test
  \citep{Esteves2016} for testing in each group
  the pragmatic Hardy-Weinberg equilibrium hypothesis
  with $m=20$. The decisions are the same for
  $KL$, $BP$ and $CD$.}
  \label{tab:hw}
 \end{table}
\end{example}

\begin{example}[Bioequivalence]
 \label{ex:bioequiv}
 Assume that $\Z=(X,Y) \sim 
 N((\mu_1,\mu_2),\sigma^2 \I_2)$, with $\sigma$ known. 
 We derive the pragmatic hypothesis for 
 $H_0:\mu_1 = \mu_2$, that is,
 for $\{(\mu_1,\mu_2) \in \Re^2: \mu_1 = \mu_2\}$.
 Such a test might be used in
 a bioequivalence study, where
 $X$ and $Y$ are the concentrations of
 an active ingredient in a generic (test) drug medication and
 in the brand-name (reference) medication \citep{chow2016analytical}, respectively.
 Since $H_0$ is composite, it is helpful to derive
 the pragmatic hypothesis of its constituents.
 
 In order to do so, let $\theta_0 = (\mu_0, \mu_0)$, 
 $\mu_0 \in \mathbb{R}$, $\theta^* = (\mu_1^*, \mu_2^*)$, and
 $H^{\theta_0} : \theta = \theta_0$.
 If $\delta_{\Z,\ts}(\z) =
 \E\left[(X-\z_1)^2 + (Y-\z_2)^2|\theta=\ts\right]$
 and $g(x)=\sqrt{x}$, then
 \begin{align*}
  \mbox{BP}_\Z(\theta_0, \theta^*)
  &=\sqrt{\frac{(\mu_1^*-\mu_0)^2+(\mu_2^*-\mu_0)^2}
  {2\sigma^2}}
 \end{align*}
 Hence, $Pg(\{\t0\}, BP_{\Z}, \epsilon) = 
 \left\{(\mu_1^*,\mu_2^*): 
 (\mu_1^*-\mu_0)^2+(\mu_2^*-\mu_0)^2 
 \leq 2\epsilon^2 \sigma^2 \right\}$
 which is a circle with center $(\mu_0,\mu_0)$
 and radius $\sqrt{2}\epsilon\sigma$, as
 depicted on the left panel of \cref{fig::bioequiv}.
 In this case, the pragmatic hypothesis is
 the Tier 1 Equivalence Test hypothesis suggested by
 the US Food and Drug Administration
 \citep{chow2016analytical}.
 The pragmatic hypothesis for $H_0: \mu_1 = \mu_2$
 is obtained by taking the union of
 the pragmatic hypotheses associated to
 its constituents, as illustrated in the right panel of
 \cref{fig::bioequiv}.
 Specifically,
 \begin{align*}
  Pg(H_0, BP_\Z, \epsilon)
  =\left\{(\mu_1^*,\mu_2^*):
  |\mu_2^*-\mu_1^*| \leq \epsilon\sigma \right\}
 \end{align*}
 
 \begin{figure}
  \centering
  \begin{subfigure}{.5\textwidth}
   \centering
   {\includegraphics[page=1,scale=.81,trim={25 20 30 30},clip]
   {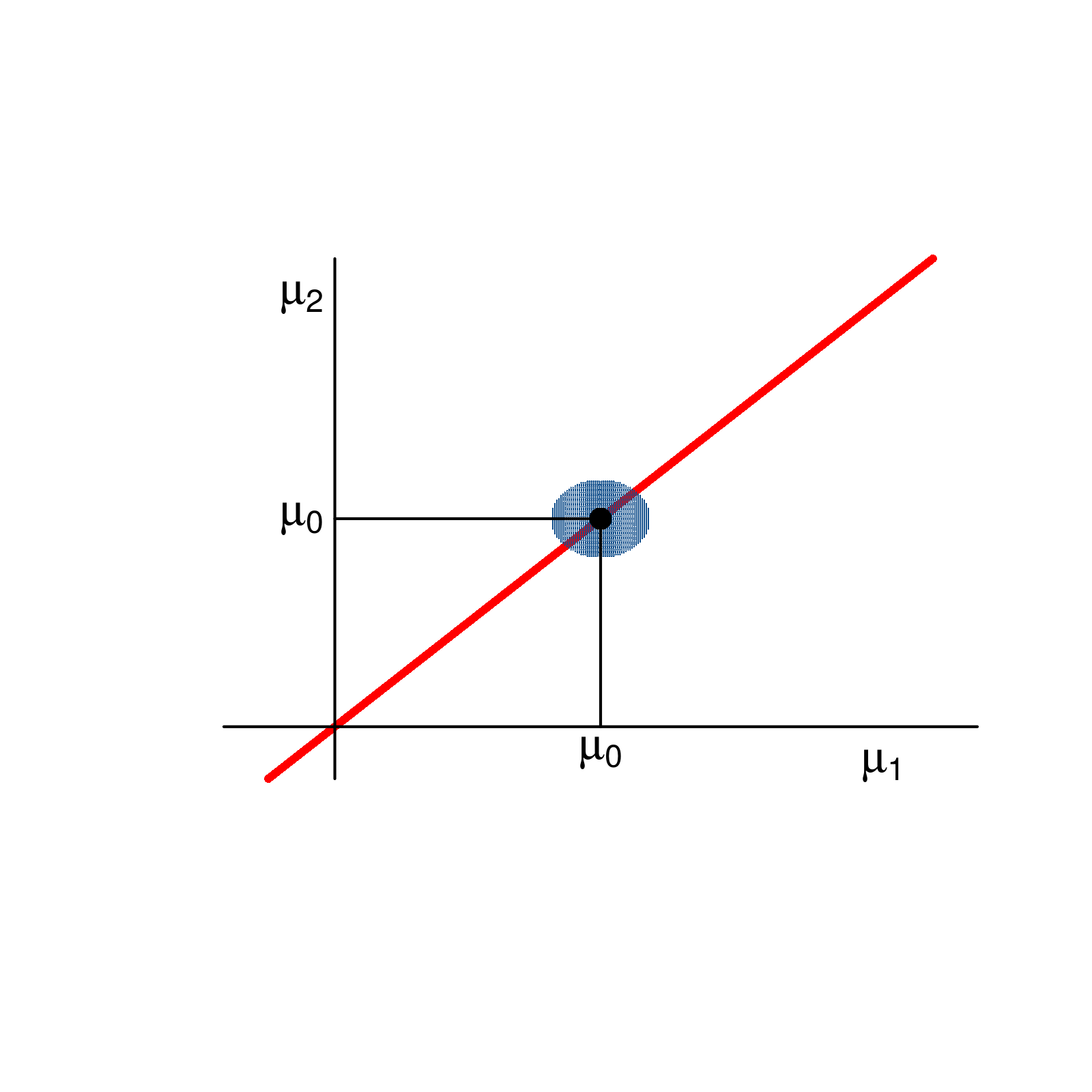}}\vspace{-1.5cm}
   \caption{$H_0: \mu_1 = \mu_2 = \mu_0$.}
  \end{subfigure}%
  \begin{subfigure}{.5\textwidth}
   \centering
   {\includegraphics[page=2,scale=0.81,trim={25 20 30 30},clip]
   {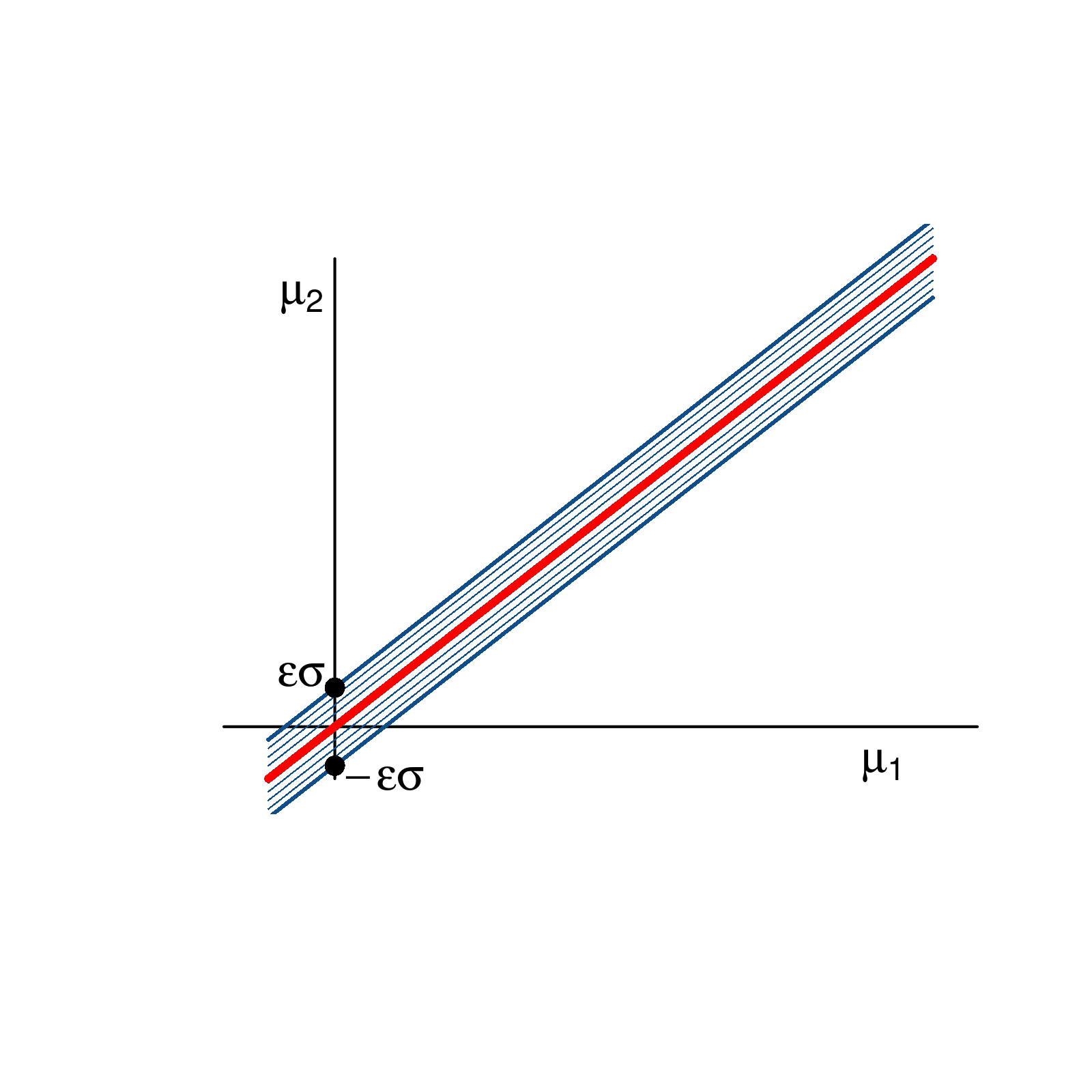}}\vspace{-1.5cm}
   \caption{$H_0: \mu_1 = \mu_2$.}
  \end{subfigure}
  \caption{Pragmatic hypotheses using BP in
  \cref{ex:bioequiv} when $\sigma$ is known.}
  \label{fig::bioequiv}
 \end{figure}
 
 The pragmatic hypothesis  for $H_0$ using KL is
 obtained similarly. Note that
 \begin{align*}
  \mbox{KL}_\Z(\theta_0,\theta^*)
  = 0.5\mbox{BP}^2_\Z(\theta_0,\theta^*)
 \end{align*}
 Therefore,
 $Pg(\{\t0\}, KL_{\Z}, \epsilon) = 
 \left\{(\mu_1^*,\mu_2^*): 
 (\mu_1^*-\mu_0)^2+(\mu_2^*-\mu_0)^2 
 \leq 2\epsilon \sigma^2 \right\}$ and
 \begin{align*}
  Pg(H_0, KL_{\Z}, \epsilon)
  &= Pg(H_0, KL_{\Z}, 0.5\epsilon^2)
 \end{align*}
 
 The pragmatic hypothesis for $H_0$ that is
 obtained using CD has no analytic expression.
 However, by observing that
 $N(\mu,\sigma^2) = \mu + \sigma N(0,1)$,
 it is possible to show that,
 there exists a monotonically increasing function,
 $h: \mathbb{R} \longrightarrow \mathbb{R}$, such that
 \begin{align*}
  Pg(H_0, CD_{\Z}, 0.5\epsilon^2)
  &= \left\{(\mu_1^*,\mu_2^*):
  |\mu_2^*-\mu_1^*| \leq h(\epsilon)\sigma \right\}
 \end{align*}
 
 That is, the pragmatic hypothesis associated to
 $H_0$ have the same shape as in
 the right panel of \cref{fig::bioequiv}.
 They differ solely on how many standard deviations
 correspond to the width of
 the pragmatic hypothesis.

\end{example}
\section{Final Remarks}
\label{sec:final}

The spiral structure studied in
\citep{Gallais1974} can be used to
describe scientific evolution.
However, in order for the analogy to
be complete, it is necessary to
indicate what types of scientific theories
or hypotheses
are effectively tested in the acceptance vertex of
the hexagon of oppositions.
We defend that these are pragmatic hypotheses,
which are sufficiently precise for
the end-user of the theory.

In order to make this statement formal,
we introduce three methods for 
constructing a pragmatic hypothesis associated to
a precise hypothesis.
These methods are based on three
predictive dissimilarity functions: KL, BP and CD.
Each of these methods have different advantages.
For instance, the scale of BP and CD is
more interpretable than KL, making it easier to
determine whether the former are large or small.
On the other hand, BP relies on the definition of
more functions than KL and CD, such as
$\delta_{\Z,\t0}(\z)$ in \cref{ex:d-pred}.
If these function are chosen inadequately, then
the shape of the resultant pragmatic hypothesis
might be counter-intuitive or meaningless.
Finally, CD often does not have an analytic expression.
It relies on numerical integration over the sample space,
which can be taxing in high dimensions.

\subsection*{Acknowledgments}

The authors are grateful for the support of IME-USP, the Institute of Mathematics and Statistics of the University of S\~{a}o Paulo, and the Department of Statistics of UFSCar - The Federal University of S\~{a}o Carlos. 
Finally, the authors are grateful for advice and comments received from anonymous referees, and from participants of the 6th World Congress on the Square of Opposition, held on November 1-5, 2018, at Chania, Crete, having as main organizers Jean-Yves B\'{e}ziau and Ioannis Vandoulakis. 
     
This work was partially supported by  \emph{CNPq -- Conselho Nacional de Desenvolvimento Cient\'ifico e Tecnol\'ogico}, grants PQ 06943-2017-4, 301206-2011-2 and 301892-2015-6; and \emph{FAPESP -- Funda\c{c}\~ao de Amparo \`a Pesquisa do Estado de S\~ao Paulo}, grants 2017/03363-8,  2014/25302-2, CEPID-2013/07375-0, and CEPID-2014/50279-4.

\bibliography{main}

\appendix

\section{Proofs}
\label{sec:proof}

\begin{proof}[Proof of \cref{thm:union}]
 Let $Pg$ be logically coherent.
 Pick an arbitrary $\t0 \in \Theta_0$ and
 note that, if $R(x) \equiv Pg(\{\t0\})$,
 then $\phi^{R}_{Pg(\{\t0\})}(x) = 0$.
 Since $Pg$ is logically coherent, conclude that
 $\phi^{R}_{Pg(\Theta_0)}(x) \equiv 0$, that is,
 $Pg(\{\t0\}) \subseteq Pg(\Theta_0)$.
 Since $\t0 \in \Theta_0$ was arbitrary, conclude that
 \begin{align}
  \label{eq:superset}
  \bigcup_{\t0 \in \Theta_0}Pg(\{\t0\}) 
  &\subseteq Pg(\Theta_0)
 \end{align}
 Next, let $R(x) \equiv 
 \bigcap_{\t0 \in \Theta_0}Pg(\{\t0\})^c$.
 For every $\t0 \in \Theta_0$,
 $\phi^{R}_{Pg(\{\t0\})}(x) = 1$.
 Since $Pg$ is logically coherent,
 $\phi^{R}_{Pg(\Theta_0)} \equiv 1$, that is,
 $Pg(\Theta_0) \subseteq R^c \equiv 
 \bigcup_{\t0 \in \Theta_0}Pg(\{\t0\})$.
 Conclude that
 \begin{align}
  \label{eq:subset}
  Pg(\Theta_0) &\subseteq
  \bigcup_{\t0 \in \Theta_0}Pg(\{\t0\})
 \end{align}
 It follows from 
 \Cref{eq:subset,eq:superset} that
 $Pg(\Theta_0) = \bigcup_{\t0 \in \Theta_0}Pg(\{\t0\})$.
 It also follows from direct calculation that,
 if $Pg(\Theta_0) = \bigcup_{\t0 \in \Theta_0}Pg(\{\t0\})$,
 then $Pg$ is logically coherent.
\end{proof}

\begin{proof}[Proof of \cref{thm:invariance}]
Let $\Theta_0 \subseteq \Theta$
 \begin{align*}
  Pg(f[\Theta_0], d_{\Z}^*, \epsilon)
  &= \{\theta^* \in \Theta^*:
  \exists \theta_0^* \in f[\Theta_0] \text{ s.t. } 
  d_{\Z}^*(\theta^*, \theta_0^*) \leq \epsilon \} \\
  &= \{\theta^* \in \Theta^*:
  \exists \theta_0^* \in f[\Theta_0] \text{ s.t. }
  d_{\Z}(f^{-1}(\theta^*), f^{-1}(\theta_0^*)) \leq \epsilon \} \\
  &= f\left[\{\theta \in \Theta:
  \exists \theta_0 \in \Theta_0 \text{ s.t. }
  d_{\Z}(\theta, \theta_0) \leq \epsilon \}\right] \\
  &= f\left[Pg(\Theta_0, d_{\Z}, \epsilon)\right]
 \end{align*}
\end{proof}

\begin{proof}[Proof of \cref{thm:convergence}]
 Since the $Z_i$'s are i.i.d.,
 $\mbox{KL}_{m}(\t0, \ts) = 
 m\mbox{KL}_{Z_1}(\t0, \ts)$.
 It follows that
 \begin{align*}
  Pg(\Theta_0, KL_m, \epsilon)
  &=\bigcup_{\t0 \in \Theta_0} 
  Pg(\{\t0\},\mbox{KL}_m,\epsilon) \\
  &= \bigcup_{\t0 \in \Theta_0} 
  Pg(\{\t0\},m\mbox{KL}_{Z_1},\epsilon)
  = \bigcup_{\t0 \in \Theta_0}
  \left\{\ts \in \Theta: 
  KL_{Z_1}(\t0,\ts) \leq m^{-1}\epsilon \right\}
 \end{align*}
 Thus,
 $\left(Pg(\Theta_0, KL_m,\epsilon)\right)_{m\geq 1}$ is
 a non-increasing sequence of sets. It follows that
 \begin{align*}
  \lim_{m \rightarrow \infty}
  Pg(\Theta_0, KL_m, \epsilon)
  &= \bigcap_{m\geq 1}\bigcup_{\t0 \in \Theta_0}
  \left\{\ts \in \Theta:
  \mbox{KL}_{Z_1}(\t0,\ts) \leq m^{-1}\epsilon \right\} \\
  &= \bigcup_{\t0 \in \Theta_0}\bigcap_{m\geq 1}
  \left\{\ts \in \Theta:
  \mbox{KL}_{Z_1}(\t0,\ts) \leq m^{-1}\epsilon \right\} \\
  &= \bigcup_{\t0 \in \Theta_0}
  \left\{\ts \in \Theta: 
  \mbox{KL}_{Z_1}(\t0,\ts) = 0 \right\} \\
   &= \bigcup_{\t0 \in \Theta_0}\{\t0\} = \Theta_0
 \end{align*}
 where the next-to-last equality follows
 from the assumption that $(F_\theta)_{\theta \in \Theta}$
 is identifiable.
 The proofs for the $CD$ divergence
 follows from the fact that
 $\mbox{TV}(\P_{\t0},\P_{\ts}) \leq \sqrt{\mbox{KL}(\P_{\t0},\P_{\ts})}$.
\end{proof}

\end{document}